\documentclass[prl,twocolumn,showpacs]{revtex4}
\usepackage{graphicx,epsfig}
\usepackage{times,amsmath,amssymb,latexsym,dsfont}
\usepackage{rotating}
\usepackage{stmaryrd}

\newcommand{\bk}[2]{\left\langle\,#1\left|\,#2\,\right\rangle\right.} %<"cosa1"|"cosa2">
\newcommand{\ket}[1]{\left|#1\right\rangle} %|"cosa">
\newcommand{\sand}[3]{\left\langle\,#1\left|\,#2\,\right|#3\,\right\rangle} %<"cosa1"|"Op"|"cosa2">
\newcommand{\ave}[1]{\left\langle\,#1\,\right\rangle} %<"Op">
\newcommand{\abs}[1]{\left|#1\right|} %|"cosa"|
\newtoks\nslashfraction\nslashfraction={.13}\newcommand{\nslash}[1]{\,\setbox0\hbox{$#1$}\setbox0\hbox to\the\nslashfraction\wd0{\hss\box0}/\box0}  %Para poder poner. Letras "slashadas" con el comando \nslash{"letra"}
\newcommand{\field}[1]{\hat\Psi(#1)}
\newcommand{\cfield}[1]{\hat\Psi^\dagger(#1)}
\newcommand{\bin}[2]{
\left( {\begin{array}{*{20}c}
   #1  \\
   #2  \\
\end{array}} \right)}
\begin{document}

\title{Vortex nucleation in  mesoscopic Bose superfluid and breaking of the parity symmetry}
\author{D. Dagnino$^{1}$, N. Barber\'an$^{1}$, and M. Lewenstein$^{2,3}$}
\affiliation{(1) Estructura i Constituents de la Mat\`{e}ria, Facultat de F\'{i}sica, Universitat de Barcelona,
E-08028 Barcelona, Spain\\
(2) ICFO - Institut de Ci\`{e}ncies Fot\`{o}niques, Parc Mediterrani de la Tecnologia, Spain\\
(3) ICREA-- Instituci\'o Catalana  de Recerca i Estudis Avan\c
cats, E-08010, Barcelona, Spain\\
}

\begin{abstract}
We analyze vortex nucleation in mezoscopic 2D Bose superfluid in a rotating trap.
We explicitly include a weakly anisotropic stirring potential, breaking thus explicitly the axial symmetry. 
As the rotation frequency passes the critical value $\Omega_c$ the system undergoes an extra symmetry change/breaking. Well below $\Omega_c$ the ground state is properly described by the mean field theory with an even  condensate wave function. Well above $\Omega_c$ the MF solution works also well, but the order parameter becomes odd. This phenomenon involves therefore a discrete parity symmetry breaking. In the critical region the MF solutions exhibit dynamical instability. The true many body state is a strongly correlated entangled state involving two macroscopically occupied modes (eigenstates of the single particle density operator). We characterize this state in various aspects: i) the eligibility for adiabatic evolution; ii) its analytical approximation given by the maximally entangled combination of two single modes; and finally iii) its appearance in particle detection measurements. 
\end{abstract}

\pacs{03.75.Hh, 03.75.Kk, 67.40.Vs}
\date{\today }
\maketitle

\section{I. Introduction}

Symmetry changes or breaking belong to some of the most fascinating phenomena in nature. In classical physics they are often associated with phase transitions in macroscopic systems
\cite{Ma,Patria}. Paradigm
examples of symmetry breaking concern magnetic
phenomena, such as for instance appearance of ferromagnets at temeperatures lower than the,  Curie temperature, $T_c$.  In the classical world  symmetry changes/breaking (C/B) are
driven by thermal fluctuations, and in the standard
Landau-Ginsburg scenario are associated with increase of classical
correlations and arousal of the long range order. Mean field approach,  that  goes back to "molecular
field theory" of Curie-Weiss \cite{Weiss:1907}, provides very often quite correct desctiption of these phenomena away from criticality. Close to critical temperature, quantitative description requires the use of renormalization group approach \`{a} la Wilson \cite{Wilson,Amitbook}.

 In quantum physics paradigm examples of
symmetry C/B deal with low temperature
behavior of weakly interacting quantum Bose gases and
Bose-Einstein condensation (BEC) \cite{pit}. In the quantum
world particularly interesting are quantum phase transitions \cite{sachdev}, and
quantum symmetry C/B that are driven by quantum
fluctuations. They  can occur either at zero temperature, or in quantum
dynamical externally driven systems.

The symmetry C/B that have drawn a lot of attention
since the early discovery of superfluids \cite{gri} till the
recent studies of BEC is nucleation of vortices in rotating
superfluids.  In fact, one of the most striking properties of superfluid and condensed systems is their response to rotation. The only way to acquire angular momentum is by the nucleation of vortices, topological singularities surrounded by condensed atoms revolving around their cores. The cores are well localized and have size of the, so called, healing length. At low temperatures they are empty, whereas at higher temperatures they are filled with  the thermal fraction of the condensate. 
For quantum
gases, atoms are usually confined in an isotropic harmonic trap and experience
an additional quadratic potential rotating at angular frequency $\Omega$ 
(for a review see \cite{fet,coo}). Standard textbooks \cite{pit} associate vortex
nucleation with thermodynamic instability. When the rotation
frequency is small, there  exists a mean field solution of the
equation describing the BEC order parameter (condensate wave
function) with a single vortex \cite{pit1,gro}; this solution
has, however, larger (free) energy than the one corresponding to
the condensate at rest \cite{strin99}. Above certain critical rotation frequency,
the solution with the vortex becomes a ground state, and in
principle may be achieved at low temperature being driven by
thermal fluctuations. In practice, experiments with BEC occur in a
completely different way. Typically, one prepares a condensate at
very low temperature, and then applies a certain dynamical
perturbation to create a vortex. First vortices have been created
at  JILA \cite{cornell} using a kind of phase imprinting method
\cite{dobrek}. It turned out, however, that the method consisting
of slight deformation and rotation of the trap, or alternatively,
"laser stirring"\cite{mad} was more efficient and led to numerous
spectacular observations such as that of Abrikosov lattice
(\cite{mad,che,mad1,ram,ram1,Haljan:2001,Hodby:2002}, for a review see \cite{fet,blo}).

Bose-Einstein condensates (BEC's) of dilute atomic gases offer particular possibilities for studies of  nucleation of vortex-states and their expansion  in the course of  time-of-flight (TOF) detection. In addition, these system, allow for experimental analysis and manipulation of mesoscopic  confined clouds of condensed atoms trapped in the sites of optical lattices. This  opens the perspective to compare directly results  of exact  numerical analysis for small systems with experiments and with different approximate calculations (for the first experiments in this direction see \cite{Chu}).
Particulalry interesting  are the studies of the applicability of the mean field (MF) theory within specific conditions. For confined systems at large rotation frequencies  (in analogy  to charged particles submitted to high magnetic fields) strongly correlated states develop, the so called fractional quantum Hall (FQH) type states \cite{Wilkin:2000,coo,Yoshioka:2002}. Such states cannot be described by a single particle wave function that would play the role of an order parameter. 

From a theoretical point of view, the
vortex nucleation can be tackled by several techniques, ranging from a MF
approach based on the Gross-Pitaevskii (GP) equation \cite{Feder:2000,sin,kas,but}
to the investigation of the many-body energy eigenstates
\cite{Bertsch:1999,Smith:2000,Jackson:2000,dag,bar,roma,parke}.
Many authors
addressed the question of vortex nucleation theoretically, asking
in the first place for energetic stability of the vortex
configuration as a ground state. Within the mean field approach
this has been discussed mainly in the context of
thermodynamic stability (cf. \cite{strin99,pit}). Several
papers discussed, however, the case of $T=0$ and vortex nucleation
in the ground state (GS) of the system using the exact  quantum
description (cf. \cite{Wilkin:2000,Smith:2000,dag,bar,mor}), or rigorous derivation
of the MF equations \cite{sei,sei1}. More recent papers treated the
problem of dynamics of vortex, or vortex lattice nucleation in
elliptically  deformed rotating traps, using mean field method
(i.e., time dependent Gross-Pitaevski equation (GPE) \cite{pit})
and trying to reproduce the experimental results. The conclusion
of these works is that vortex nucleation is inevitably associated
to dynamical instability of the solutions of GPE  \cite{sin,ros}.
Same results hold in the case of vortex nucleation via phase
imprinting \cite{andrelczyk}. Some authors \cite{kas}
    claim that apart from the  dynamical instability, a Landau instability
(associated to dissipation) is also necessary to allow vortices to
penetrate the BEC. The dynamical instability GPE is generically
associated to the appearance of squeezing of two Bogoliubov-de
Gennes (BdG) quasi-particle modes, i.e., exponentially growing two
mode entanglement in the regime of validity of BdG (i.e., regime of
small Gaussian fluctuations around the MF solutions) \cite{garay}.
This observation already indicates the necessity of going beyond
the mean field at the instability. 

Much less is known about exact dynamics of the
vortex nucleation. Parke {\it et al.} \cite{parke} considered
recently this problem for a mesoscopic sample of atoms in the lowest
Landau level (LLL) and discover striking non-mean field effects in
the (stationary) spectrum of the system at the critical rotation
frequency $\Omega_c$. They interpreted their results in terms of
cooperative tunnelling of a vortex pair by "requantizing" the mean
field theory (reduced to 3 relevant modes). In a recent paper \cite{nature} we used another approach 
 and studied exact dynamics of a mesocopic sample of atoms in
a elliptically deformed rotating harmonic trap. Our main result was
that as one increases $\Omega$, at  the  $\Omega_c$ the mean field
description ceases to be valid. The system enters a strongly
correlated and entangled state, well described by an effective two
mode model. The mean field description (similar to that of ref.
\cite{parke}) exhibits dynamical instability and hysteresis for
$\Omega\simeq\Omega_c$. Since we explicitly include an anisotropic stirring potential, 
the present mechanism concerns  a discrete parity symmetry breaking. Therefore it 
differs from the case of the vortex nucleation in axially symmetric traps: 
in the latter case, breaking of the continuous rotational symmetry involves 
a gapless Nambu-Goldstone mode \cite{Ueda}, while here we deal with a gapped system. 

We believe that this example constitutes a
paradigm of mean field symmetry C/B in the course of
adiabatic evolution of a many-body system. The character of the strongly 
correlated states depend on the nature and character of the symmetry C/B, and the specific system - it is thus different 
in our case and in the case of Ueda {\it et al.} \cite{Ueda}, or in the case of rotating lattice rings \cite{nun}, or in BEC in a tilted double well potential \cite{chweiss}. The last reference \cite{chweiss} reveals however, a feature closely related to our results, which may have universal character: instability or chaotic behaviour of the system predicted by the MF approach, is a signature of the existence of a strongly correlated state.

This is quite an unexpected result, since at least for large systems, when $N$ goes to infinity,  the MF theory in the regime of nucleation of the first vortex where the angular momentum of the systems changes from $L=0$ to $N$ is believed to be correct. This belief has been in fact recently supported by rigorous results in Ref.\cite{sei1}, where it has been proved that the MF expressions for the total enerergy of the system coincides with the exact result for large $N$. Moreover, while the MF description of the  GS is not correct for moderately big $N$ at criticality, the single particle density is another example of a quantity correctly described by the MF approach. This last observation is supported by our result: we obtain that the density of the ground state at $\Omega_c$ for increasing $N$ becomes indistinguishable from that obtained for small $N$. This means that some 
macroscopic mean quantities, like the density or the total energy, are practically insensitive to the symmetry C/B process at criticality.

The aim of the present work is to perform a deeper analysis of the precursor state of the nucleation process, named $\Psi_c$ from now on,  using various techniques. First, we analyze the dynamical process of increasing rotation frequency and show that, during the time evolution from an initial GS at a starting rotation frequency (where the angular momentum is $L=0$) to the final one-vortex state where $L=N$, one must pass through a state that cannot be described by a MF  order parameter. This state  contains two  macro-occupied modes of different parity of the single particle density matrix.   Second, we characterize this non-condensed state and its properties, ask how does this state exhibit these properties in measurement process. We note that similar
questions have been posed in the context of appearance of the
relative phase in the interference of two BECs  in the course of
measurements \cite{jav,cast,mul}. Inspired by the  work of Javanainen et al. \cite{jav} we simulate the measurement process in a TOF experiment, for a one shot event, and for the acumulation of a large number of single shots. To be able to perform the measurement simulation of the $\Psi_c$ state assuming a large number of particles, we use a two-mode model which provides a very  accurate approximation of the exact state. Most of the times, the single shot events produce a single vortex located randomly along the $x$-axis whereas the accumulation of a large number of shots, reproduce the density. We discuss the interpretation  of the outcomes and relate them with other strongly correlated states, discussed previously in various systems. 

Our paper is organized as follows: In Section II we present our model, and a brief repetition of previously obtained  results for rotating bosonic systems. In Section III  we analyze the time evolution of the nucleation process. First we look at the possibility of an  adiabatic evolution,  and secondly, we discuss a two-mode model for the GS at a critical frequency at which parity symmetry-breaking takes place at the MF level. In Section IV we study the energy spectrum as a function of the rotation frequency, in terms of the contributions of different $L$-subspaces in the GS and analyze the robustness of the $\Psi_c$ state. In Section V we describe measurement simulations and discuss their possible interpretations. Here the comparison with a model cat-state  (analogous to that predicted in Ref. \cite{nun}) is included. Finally in Section VI we present our conclusions.

\section{II. Model and background}
We assume a two-dimensional cloud of few condensed Bose atoms of mass $M$ interacting via contact forces, confined in a symmetric parabolic trap with frequency $\omega_{\perp}$. Two extra perturbing potentials are also considered: One simulates a stirring laser that sets the system in rotation by a slight anisotropic deformation in the $xy$ plane, rotating at angular frequency $\Omega$ around the $z$-axis breaking the cilindrical symmetry. The second one, is a perturbation that breaks the parity symmetry, a symmetry that is otherwise preserved by the previous terms. The last one simulates possible second order contributions of the laser fields, and will help us in the analysis of the system. The rotation frequency $\Omega$ is strong enough to assume the lowest Landau level (LLL) regime \cite{mor}, i.e., we consider that the kinetic energy within the LLL is  given by $\hbar(\omega_{\perp}-\Omega)$, the strength of the interaction, and both perturbations, are small compared with the separation between Landau levels given by $\hbar(\omega_{\perp}+\Omega)$ \cite{jacak}. It is implicit in our model that in the $z$ direction, a strong parabolic trap of frequency $\omega_z$ freezes the atomic motion producing an effective two-dimensional (2D) system. We model the contact interaction $U$, and the two perturbing potentials $V_1$ and $V_2$ by,
\begin{equation}
U(\vec{r_i},\vec{r_j})=(\hbar^2 g/M)\sum_{i<j}^N \delta(\vec{r_i}-\vec{r_j})
\end{equation}
\begin{equation}
V_1(\vec{r})=2AM\omega_{\perp}^2\sum_i^N(x_i^2-y_i^2)
\end{equation}
and
\begin{equation}
V_2(x)=B\frac{M\omega_{\perp}^2}{\lambda_{\perp}}\sum_i^N x_i^3
\end{equation}
where $g=\sqrt{8\pi}a/\lambda_z$, $a$ being the 3D scattering length, $\lambda_z=\sqrt{\hbar/M\omega_z}$ and $\lambda_{\perp}=\sqrt{\hbar/M\omega_{\perp}}$. The dimensionless parameters $g$, $A$ and $B$ measure the strength of each term. We choose $\lambda_{\perp}$, $\hbar \omega_{\perp}$ and $\omega_{\perp}$ as units of length, energy and frequency respectively. It is worth mentioning that a simpler term that breaks the parity symmetry $\sim B\sum_i x_i$ is a center of mass excitation that would leave the internal structure unchanged revealing no new physics. In the second quantized
formalism the Hamiltonian of the system projected onto the LLL in the rotating reference frame is described by:

\begin{equation}
\hat{H}= \alpha \hat{L} + \beta \hat{N} + \hat{U} +
\hat{V_1} + \hat{V_2}\,\,\equiv \,\,\hat{H_0} + \hat{U} + \hat{V_1}+\hat{V_2}\,\,,
\label{secondquant}
\end{equation}
where $\alpha=\hbar(\omega_{\perp}-\Omega)\,\,$ and $\beta=\hbar\omega_{\perp}$. $\hbar\hat{L}$ , and $\hat{N}$ are the total $z$-component angular momentum and particle number operators, respectively, $H_0$ being the kinetic contribution. The contact interaction term is given by the operator

\begin{equation}
\hat{U}=\frac{1}{2}\sum_{m_1m_2m_3m_4}
U_{1234}\,\,\,a^\dag_{m_1}a^\dag_{m_2}a_{m_4}\,a_{m_3}\,\,\,\,,
\end{equation}
where the matrix elements read

\begin{eqnarray}
U_{1234}& = &\langle m_1\,m_2 \mid U \mid m_3\,m_4 \rangle
\nonumber
\\
& = &\frac{g}{\lambda_{\perp}^2 \pi}\,\,\frac{\delta_{m_1+m_2,
m_3+m_4}}{\sqrt{m_1!m_2!m_3!m_4!}}\,\,
\frac{(m_1+m_2)!}{2^{m_1+m_2+1}}\: .
\end{eqnarray}
The operators $a^{+}_{m_i}$ and $a_{m_i}$ create and annihilate a boson in a single particle eigenstate of $\hat{H_0}$ with angular momentum $m_i$, respectively. These eigenstates are taken as a basis to represent wavefunctions and operators in the second quantized formalism. We will refer to this set of functions as the Fock-Darwin (FD) functions restricted to the LLL,  (without nodes in the radial direction) and will denote them as $\varphi_m=\frac{1}{\sqrt{\pi m!}}(x+iy)^m e^{-(x^2+y^2)/2\lambda_{\perp}^2}$ being $m$ its angular momentum. The term $\hat{V_1}$ in Eq.4 is given by:

\begin{eqnarray}
\hat{V_1}& = &\frac{A}{2}\sum_m
\sqrt{m(m-1)}\,\,a^{\dag}_ma_{m-2}
\nonumber
\\
& + &\sqrt{(m+1)(m+2)}\,\,a^{\dag}_ma_{m+2}\,\,.
\end{eqnarray}
and the term $\hat{V_2}$ by,

\begin{eqnarray}
\hat{V_2}& = &\frac{B}{8}\sum_m(\sqrt{m(m-1)(m-2)}\,\,a^{\dag}_ma_{m-3}
\nonumber
\\
& + &3(m+1)\sqrt{m}\,\,a^{\dag}_ma_{m-1}
\nonumber
\\
& + &3(m+2)\sqrt{m+1}\,\,a^{\dag}_ma_{m+1}
\nonumber
\\
& + &\sqrt{(m+1)(m+2)(m+3)}\,\,a^{\dag}_ma_{m+3})\,\,.
\end{eqnarray}

In the absence of anisotropy ($A=B=0$) the total angular momentum of the GS as a function of $\Omega$ shows sharp steps at critical values $\Omega_i$ $i=1,2...$, being $\Omega_1$ ($\Omega_1=\omega_{\perp}-gN/8 \pi$) the value at which the angular momentum of the system jumps from zero to $L=N$ for all $N$. Above $\Omega_1$ a plateau indicating constant angular momentum extends up to $\Omega_2$ where the second jump, not always to $L=2N$ takes place. From this value, a sequence of jumps and plateaux emerge up to the last possible $L$-value given by $L=N(N-1)$, corresponding to the Laughlin state. The extension of the first plateau, from $\Omega_1$ to $\Omega_2$ increases with $N$ and at the same time, $\Omega_1$ decreases (this is a characteristic of the first plateau, in contrast, the next ones reduce drastically as $N$ increases). From the expression of $\Omega_1$ it is evident that for a given value of $g$ there is a maximum number of atoms compatible with our LLL assumption; for large $N$ ($N\geq 25$ for $g=1$) the functional relation between $\Omega_1$ and $N$ cannot be linear any more, extra Landau levels must be taken into account.

Eigenstates with more than one vortex can only exist locally at $\Omega_i$ where several eigenfunctions with different angular momentum, but degenerated in energy can be combined to generate vortex configurations in a spontaneous symmetry breaking mechanism. For instance, for $N=5$ there is degeneracy between the states with $L=0,2,3,4$ and $5$ at  $\Omega_1$, or $L=5$ and $8$ at $\Omega_2$ or $L=8,10$ and $12$ at $\Omega_3$. Namely, for small systems, the posibility to have vortex states is localized around discrete values of $\Omega$. However for larger $N$, on the one hand, the distance between the critical values of $\Omega_i$  is drastically reduced \cite{bar}, and on the other hand, the steps are softened due to the slight anisotropy that must be applied to accelerate the system; these two effects effectively provide the possibility to nucleate vortices in a continiuous way for all $\Omega>\Omega_1$ \cite{che,sei}. 

If anisotropy is included in the Hamiltonian, the steps are soften. For a fixed $A$ (and $B=0$) the softening is larger for small values of $N$ in such a way that the steps may disappear. This effect can be reduced increasing $g$, namely, the increase of anisotropy reduces the effect of the interaction. To keep a well defined first step around the same $\Omega_1$ for all $N$ we considered $gN=$$constant$ in our calculations; in addition, this decrease of the interaction as $N$ increases preserves the LLL condition \cite{mor}. 

Having specified the Hamiltonian of the model, we perform exact diagonalization. The isotropic Hamiltonian can be diagonalized in boxes of finite L-subspaces whereas, in the anisotropic case, several basis of different L-subspaces must be considered untill convergence is obtained, depending on the value of $A$ and $B$. However, if $B=0$ since the term $\hat{V_1}$ in Eq. (7) can only mix $L$ and $L\pm 2$, the parity of the angular momentum is well defined and only subspaces with the same parity must be considered.

To finish this section, we must introduce the single particle density matrix (SPDM) operator  and its eigenfunctions as tools to be used in the next sections. One of our aims is to analyse the states generated as $\Omega$ grows in a dynamical process that follows the evolution of the GS from an initial value $\Omega_{i}$ (smaller than  a critical frequency $\Omega_c$ to be defined later) to a final value $\Omega_f>\Omega_c$ at the middle of the first plateau, where the first vortex has been nucleated. For these relatively low values of $\Omega$ and far from the critical frequency, the degree of condensation is high and the eigenfunction
(a single particle wave function) of the SPDM that corresponds to the highest occupation plays the role of the order parameter (OP) of the condensate \cite{dag2}. To obtain this OP we solve the eigenvalue equation for the SPDM, 

\begin{equation}
\int d\vec{r'} n^{(1)}(\vec{r},\vec{r'})
\psi^{*}_k(\vec{r'})= n_k
\psi_k(\vec{r}),
\end{equation}
where

\begin{equation}
n^{(1)}(\vec{r},\vec{r'}) = \langle GS\mid \hat{\Psi}^{\dag}(\vec{r})
\hat{\Psi}(\vec{r'})|GS \rangle,
\end{equation}
with $\hat{\Psi}$ being the field operator. If there exist a relevant eigenvalue $n_1>>n_k$ for $k=2,3,...,$ then

\begin{equation}
\sqrt{n_1} \psi_1(\vec{r})
\end{equation}
plays the role of the OP of the system. The map of the local phase of this complex function gives precise information of the position of vortices \cite{dag2}. The change of the phase by $2\pi \nu$ around a point, where $\nu$ is a positive interger signals the location of a vortex with $\nu$ quanta of circulation. Notice that $m$ labels the single-particle angular momentum from $m=0,1,2,...$ of the FD functions, whereas $k=1,2,...$ is a label that distinguishes between the eigenstates of the SPDM. In Appendix $A$ we show a detail derivation of the relationship between the operators that create FD functions: $\hat{a_m}^{\dag}$ and those that create eigenfunctions of the SPDM operator: $\hat{b_k}^{\dag}$. 
\medskip

\section{III. Dynamics of the nucleation of the first vortex}

\subsection{Adiabatic evolution}

First, we try to find out  if the nucleation of the first vortex can be obtained as the final configuration  of the adiabatic time evolution of the initial GS submitted to increasing $\Omega$. To understand  how the resulting state changes during the process, we perform several runs from $\Omega_0$ to $\Omega_f$ for increasing time intervals and compare the results with the sequence of stationary solutions obtained from the diagonalization of the time-independent Hamiltonian, for the same range of $\Omega$ values. 

Assuming linear dependence in time of the rotation frequency as
\begin{equation}
    \Omega(t)= \Omega_{i} + \gamma t
\end{equation}
where $\gamma$ is a constant,
we solve the Schr\"odinger equation
\begin{equation}
	i\partial_t\ket{\Phi(t)}=\hat{H}(t)\ket{\Phi(t)}
\end{equation}
for the initial exact solution $\ket{\Phi(t=0)}$ at $\Omega_{0}$.
If we assume the expansion of  the state in the Fock basis as
$\ket{\Phi(t)}=\sum_kc_k(t)\ket{k}$ where $\ket{k}$ is the $N$-particle Fock state with the well defined angular momentum, and project Eq.(13) on the state $\ket{j}$, we obtain the system of equations,
\begin{equation}
	i\partial_tc_j(t)=\sum_{k}c_k(t)\sand{j}{\hat{H}(t)}{k}\,\,\,,
\end{equation}
which can be solved numerically. The bracket
$\sand{j}{\hat{H}(t)}{k}$ is expressed as
\begin{equation}
	\sand{j}{\hat{H}(t)}{k}=\sand{j}{\hat{H}_1}{k}+L_j(1-\Omega(t))\delta_{jk},
	\label{eq:evo}
\end{equation}
with a time-independent matrix which must be calculated only once, plus a time-dependent diagonal term. $\hat{H}_1=\beta \hat{N}+\hat{U}+\hat{V_1}+\hat{V_2}$; $L_j$ is here the angular momentum of the N-particle Fock state $\ket{j}$.
\begin{figure}[tbp]
	\centering
		\includegraphics*[width=1.0\columnwidth]{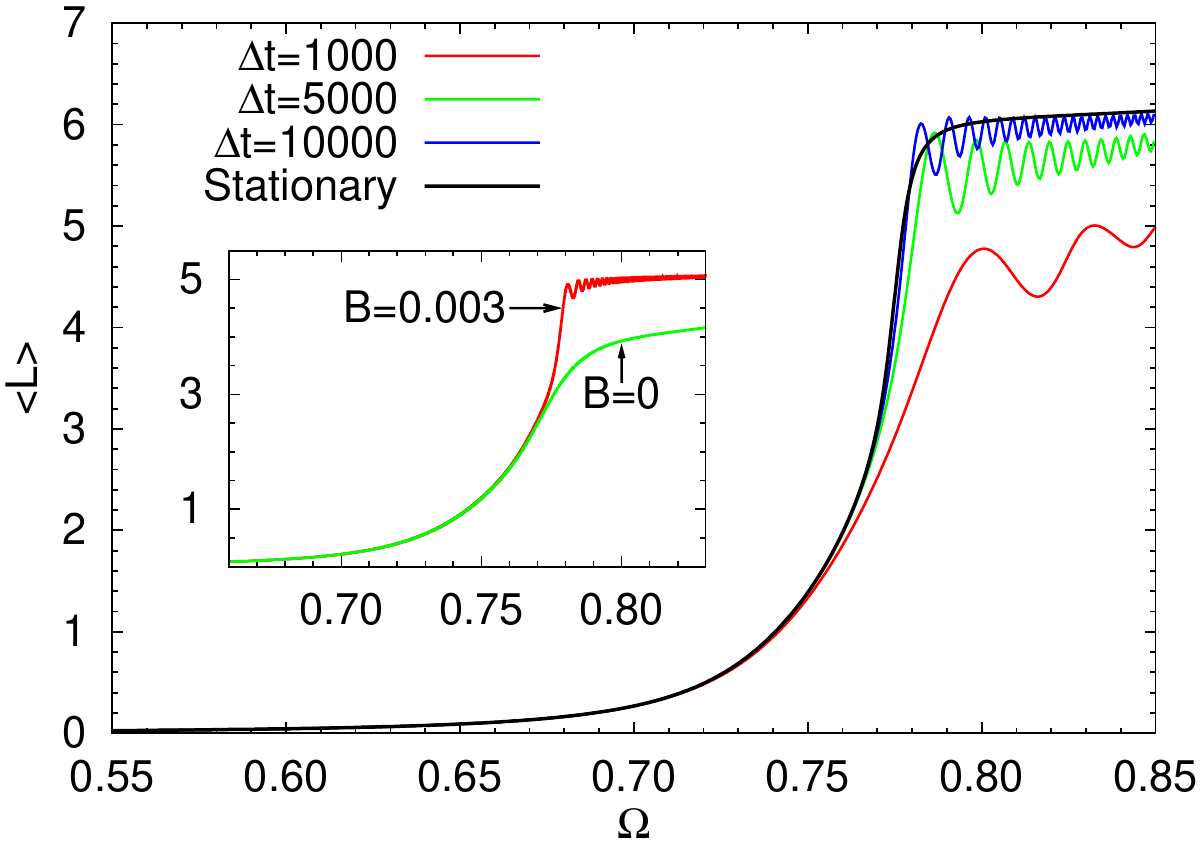}
\caption{For $N=6$ and $B=0$, time evolution of the mean value of the angular momentum, taking as the initial condition its GS-value at $\Omega_0=0.4$ (not shown in the figure). The value of $\Delta t$ is the time used in the process to run linearly from $\Omega_{0}$ to $\Omega_f=0.85$ in units of $\omega_{\perp}^{-1}$. For $N=5$, the inset shows $<L>$ over $\Omega$; for $B=0$ once adiabaticity is fulfilled and close to  adiabaticity for $B=0.003$. $\Delta t=50000$ in the last case. ($A=0.03$ and $g=1$ has been considered).}
\end{figure}

We compare then the time evolution with the corresponding sequence of stationary states at instant rotation frequencies, and choose three different criteria of adiabaticity. The first one compares the profiles of the angular momentum as a function of $\Omega$. We consider that adiabaticity is obtained when the maximum difference between the curves is $0.03$. In Fig. 1 we plot for $N=6$ ($A=0.03$ and $B=0$) the evolution of the angular momentum of the state that fulfills the condition of being the exact GS at $\Omega_0=0.4$, up to $\Omega_f=0.85$ at the middle of the first plateau. Different time intervals $\Delta t$ have been considered untill adiabaticity is achieved for $\Delta t=30000$; the black line corresponds to the stationary solutions. This means that from the stationary state at $\Omega_0=0.4$ a slow evolution taking at least $28$ seconds is necessary to nucleate the first vortex at $\Omega_f$ evolving through stationary states ($g=1$, $A=0.03$ and $\omega_{\perp}=2\pi\times 175$ Hz has been considered). The equilibrium state at $\Omega_0=0.4$ can be experimentally realized after relaxation, once the system is suddenly put in rotation at $\Omega=0.4$ \cite{mad}. The origin of the oscillations can be clarified by the analysis of the contributions of different $L$-subspaces in the expansion of $\ket{\Phi(t)}$ as $\Omega$ evolves. Fig. 2 shows the weight of each subspace within $\Phi(t)$, it can be inferred that the interplay between $L=6$ and $L=4$ produces the oscillations. We proved that this is a general result for all $N$, very fast evolution is possible, keeping adiabaticity, from $\Omega_0$ up to about $\Omega_1$, whereas beyond this frequency, a sequence of quadrupolar excitations  between $L=N$ and $N-2$ produce oscillations on $<L>$ lowering the speed.              
    
\begin{figure}[tbp]
	\centering
		\includegraphics*[width=1.0\columnwidth]{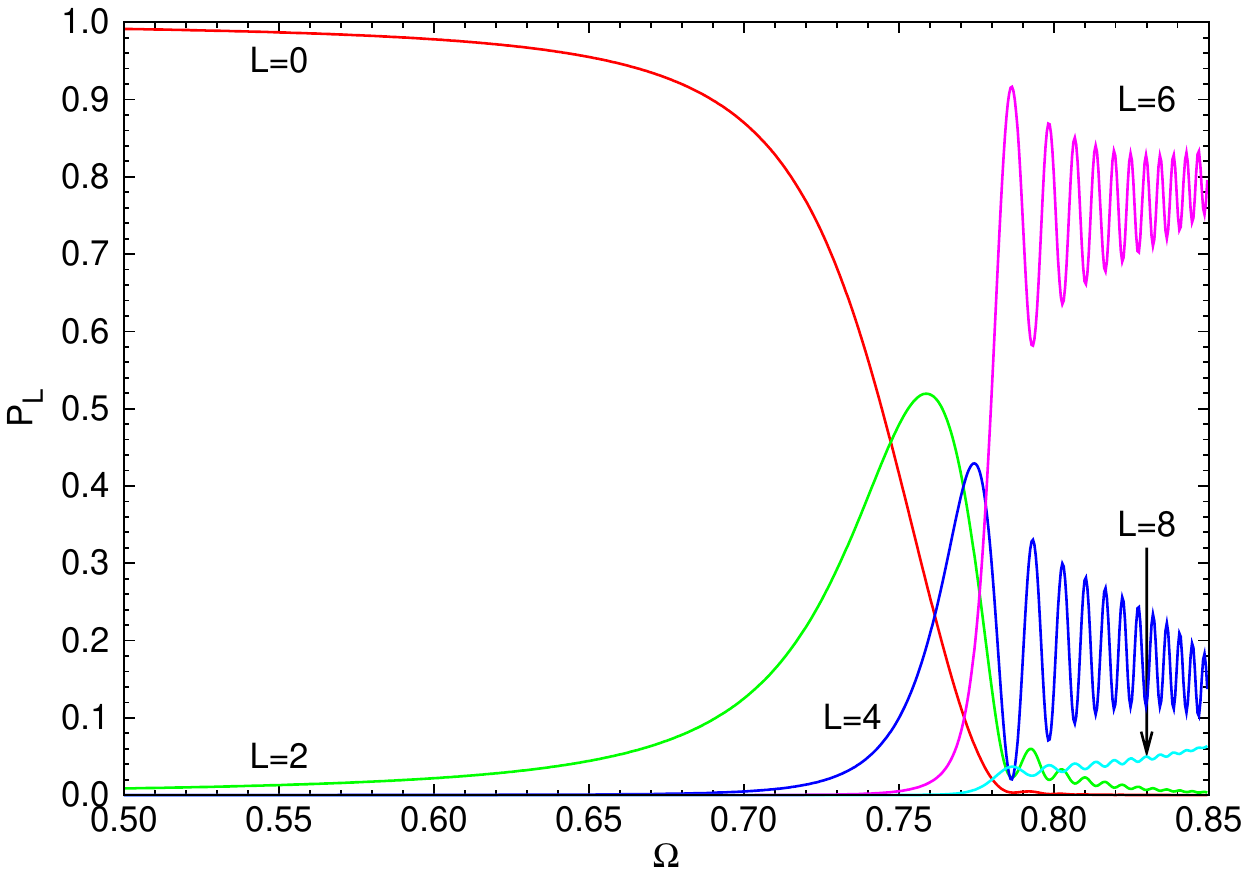}
	\caption{Evolution of the weight ($P_L$) of the $L$-subspace components within $\Phi(t)$ before adiabaticity, for $N=6$ $g=1$, $A=0.03$ and $B=0$.}
\end{figure}

A similar behaviour is found for our second criterion based on the evolution of the expected value of the energy. In contrast to the previous case however, the coincidence with the stationary values is obtained much faster, (about $\Delta t=5000$ for the GS),  producing the incorrect impression of adiabaticity, incorrect since other characteristics of the system, as is the case of the angular momentum,  are still not reproduced. 
 
Finally, the third option is to  measure the overlap between the exact GS and $\ket{\Phi(t)}$ at  $\Omega_f$ and to consider that one gets adiabaticity when the overlap is larger than $0.99$. For $N=6$ adiabaticity is fulfilled in $21$ sec, compatible with the result obtained from the first criterion.

For odd values of $N$ the time evolution driven by a parity-preserving Hamiltonian (with $B=0$) cannot carry up $<L>$ from $0$ to $N$, and a slight perturbation, which breaks parity symmetry is necessary. In other words, the sequence of stationary solutions of the parity invariant Hamiltonian develop a first order transition at about $\Omega_1$ where even $L$-subspace contributions are substituted by the odd ones within the composition of the GS. In the inset of Fig. 1 we show, for comparison, for $N=5$ the time evolution for the case when  adiabaticity is achieved for $B=0$,  and nearly achieved for $B=0.003$ ($\Delta t=50000$ in the last case). In the first case, the expected value of the angular momentum is around $4$, and no vortex is nucleated, whereas in the second case a one-vortex state with $<L>=5$ is obtained.

Figure 3 shows $\gamma_c$ defined as $\Omega_f=\Omega_0+\gamma_c \Delta t$ as a function of $N$. For fixed values of the initial and final frequencies taken for all cases at $\Omega_0=0.4$ and $\Omega_f=0.85$, we increase $\Delta t$ until adiabaticity is fulfilled (using the third criterion previously mentioned), and from it, we obtain $\gamma_c$. In both cases, with and without the parity-breaking term in the Hamiltonian, $\gamma_c$ converges, meaning that even for large number of particles, the process is possible at a finite time interval.

\medskip

\begin{figure}[tbp]
	\centering
		\includegraphics*[width=1.0\columnwidth]{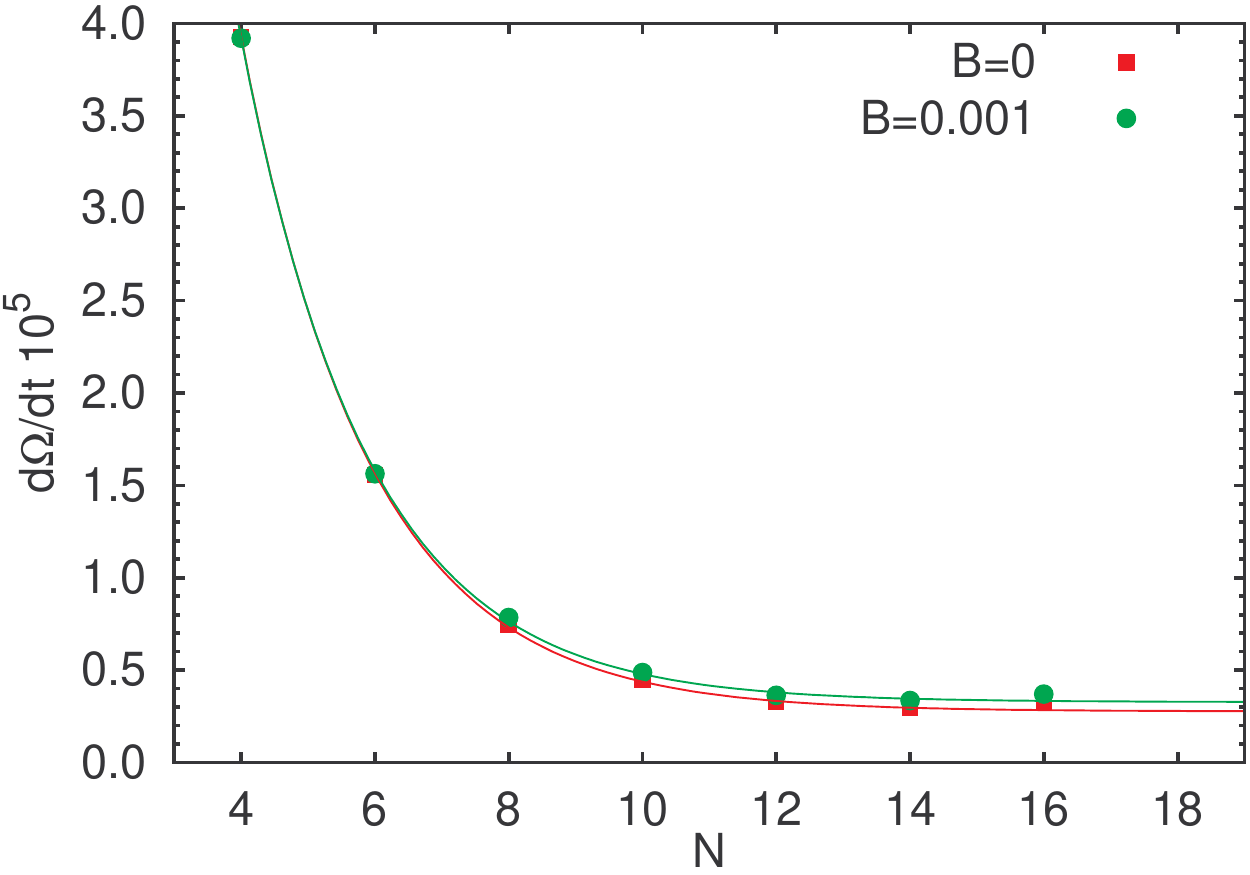}
	\caption{ Dependence on $N$ of $\gamma_c$, the critical value of the variation of $\Omega$ with time, necessary to arrive to adiabaticity. $B=0$ and $B=0.001$ has been considered for the upper and lower curves respectively, and $A=0.01$ and $gN=6$ in both cases. The horizontal line marks the asymptotic value for large number of atoms.}
\end{figure}

\subsection{The two mode state}

\begin{figure}[tbp]
	\centering
		\includegraphics*[width=1.0\columnwidth]{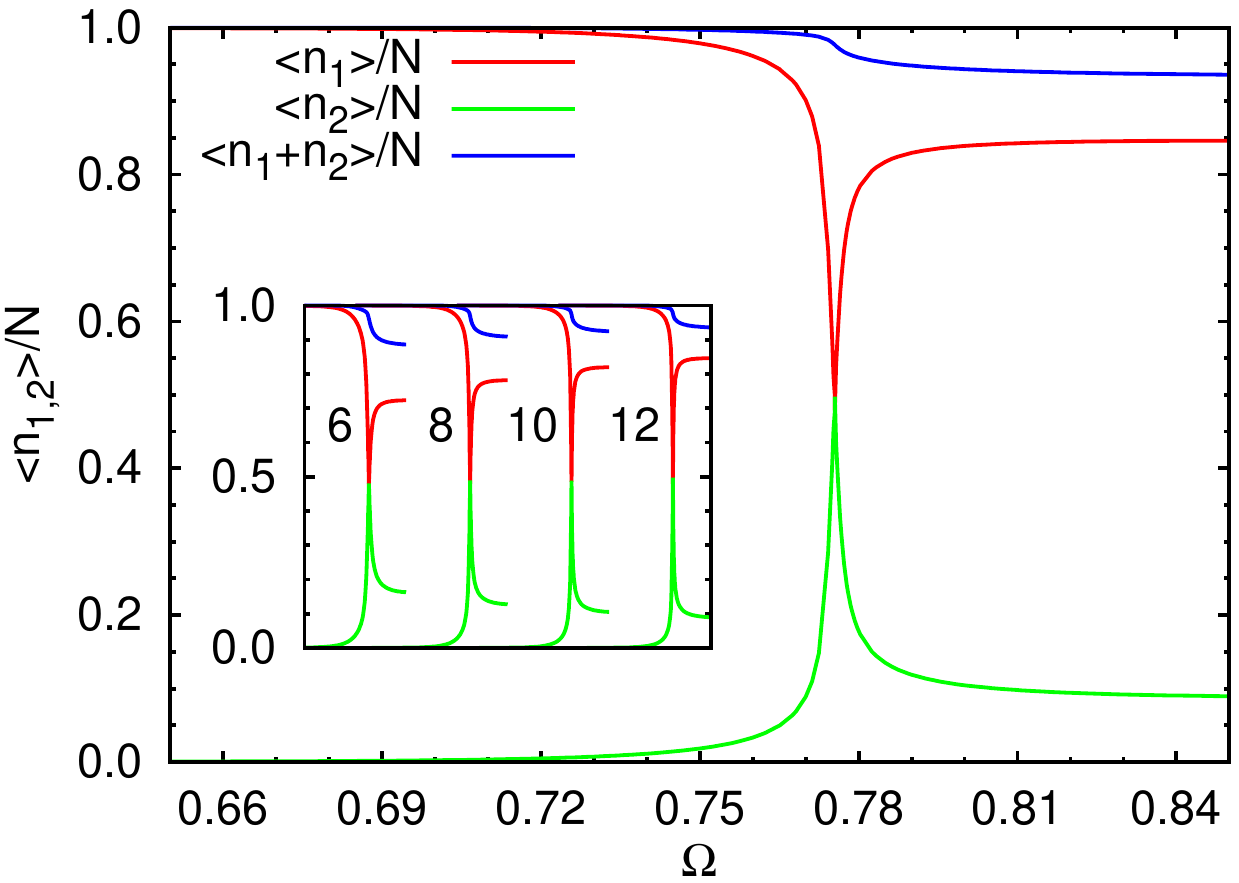}
	\caption{ Evolution of the mean value of the occupations $n_1/N$ and $n_2/N$ in the GS, as a function of $\Omega$, once adiabaticity is fulfilled for $N=12$. The peaks touch at $\Omega_c= 0.7759$. The addition of both contributions is also shown. The inset shows the results for different number of particles. As $N$ increases, the condensation far from $\Omega_c$ increases, and at the critical frequency, the value of $n_{1,2}$ is closer to $0.5$ meaning that the two mode-model is indeed better. The graphs are horizontaly shifted for clarity . $A=0.03$ and $gN=6$ has been considered.}
\end{figure}

\begin{figure}[tbp]
	\centering
		\includegraphics*[width=1.0\columnwidth]{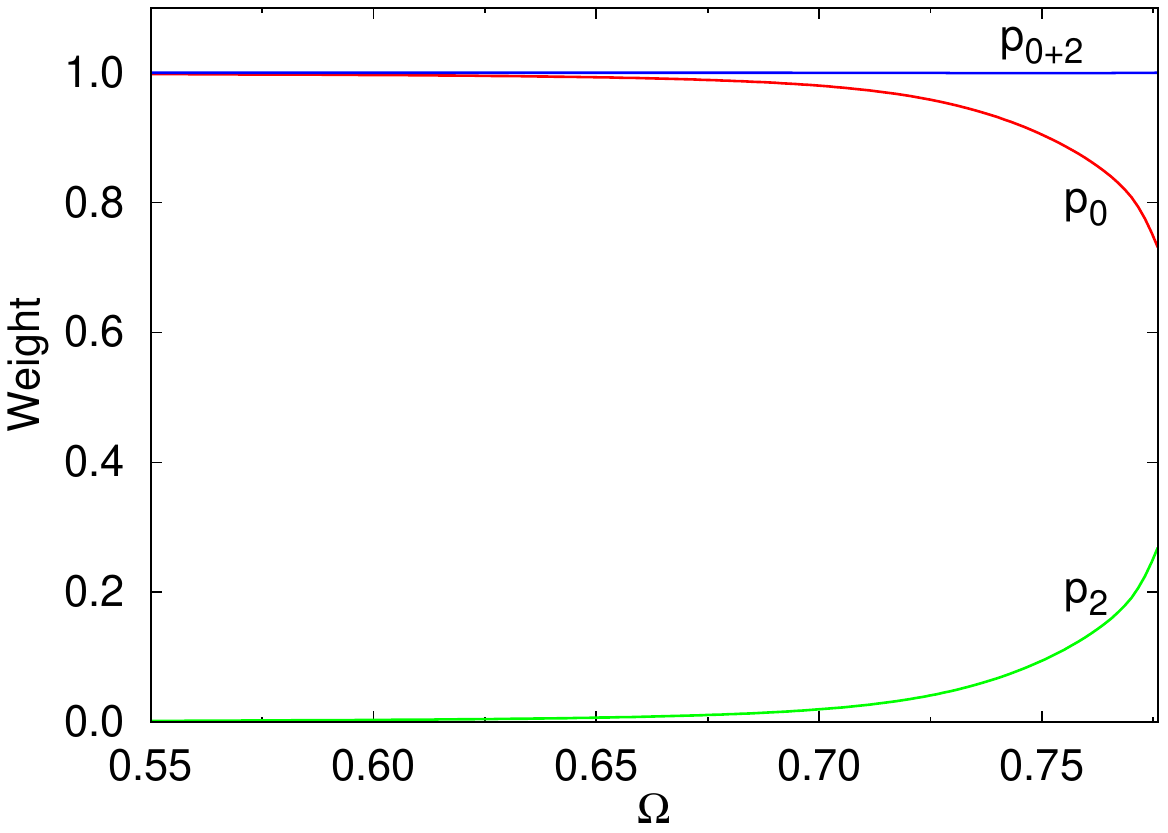}
	\caption{ Weight versus $\Omega$ of the FD functions $\varphi_0$ and $\varphi_2$ ($p_0$ and $p_2$) in the expansion of the most occupied wave function $\psi_1$, below $\Omega_c$ as well as their joint contribution. $N=12$, $A=0.03$ and $gN=6$ has been considered.}
\end{figure}

\begin{figure}[tbp]
	\centering
		\includegraphics*[width=1.0\columnwidth]{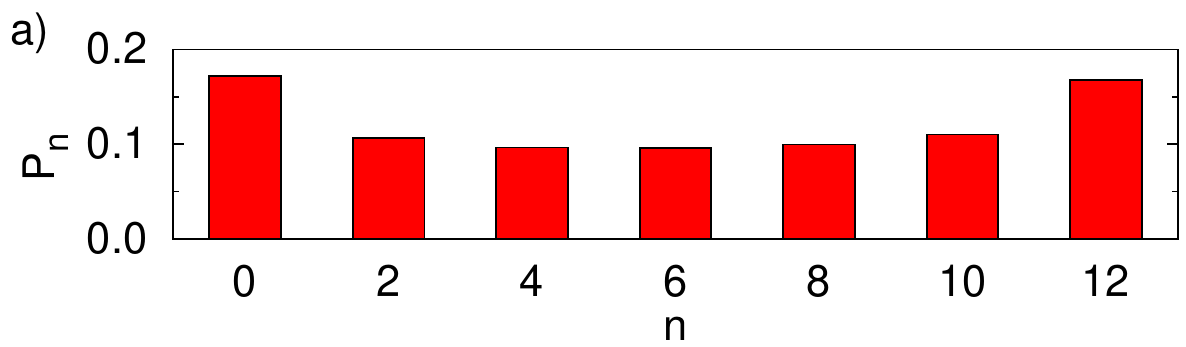}
    \includegraphics*[width=1.0\columnwidth]{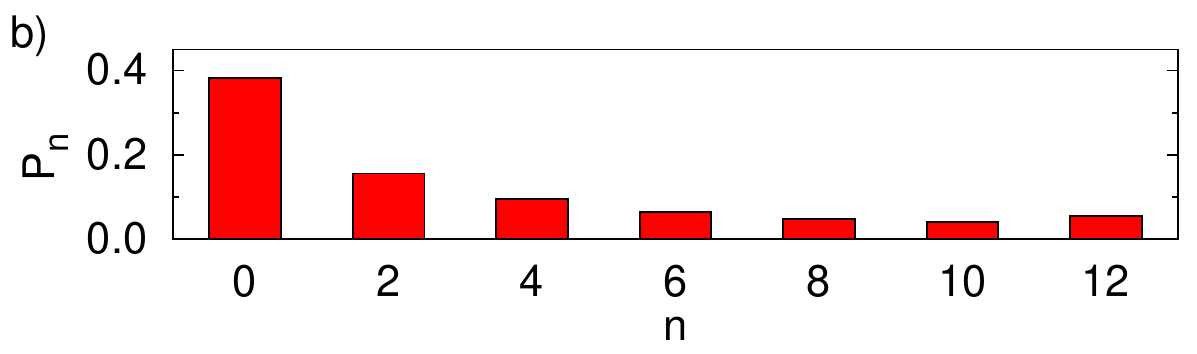}
	\caption{ Analysis of the GS for $N=12$ at the critical frequency in terms of the square of the scalar products, $P_n=\mid \langle n:\psi_1\; ;\; N-n:\psi_2|\Psi_0\rangle \mid ^2$. Fig. 6a is for $B=0$ and 6b for $B=0.001$. $\mid \langle ME \mid GS \rangle \mid=0.92$ for $B=0$ and $\mid \langle PB \mid GS \rangle \mid=0.840$ for $B=0.001$. $A=0.03$ and $gN=6$ in both cases.}
\end{figure}

\begin{figure}[tbp]
	\centering
		\includegraphics*[width=1.0\columnwidth]{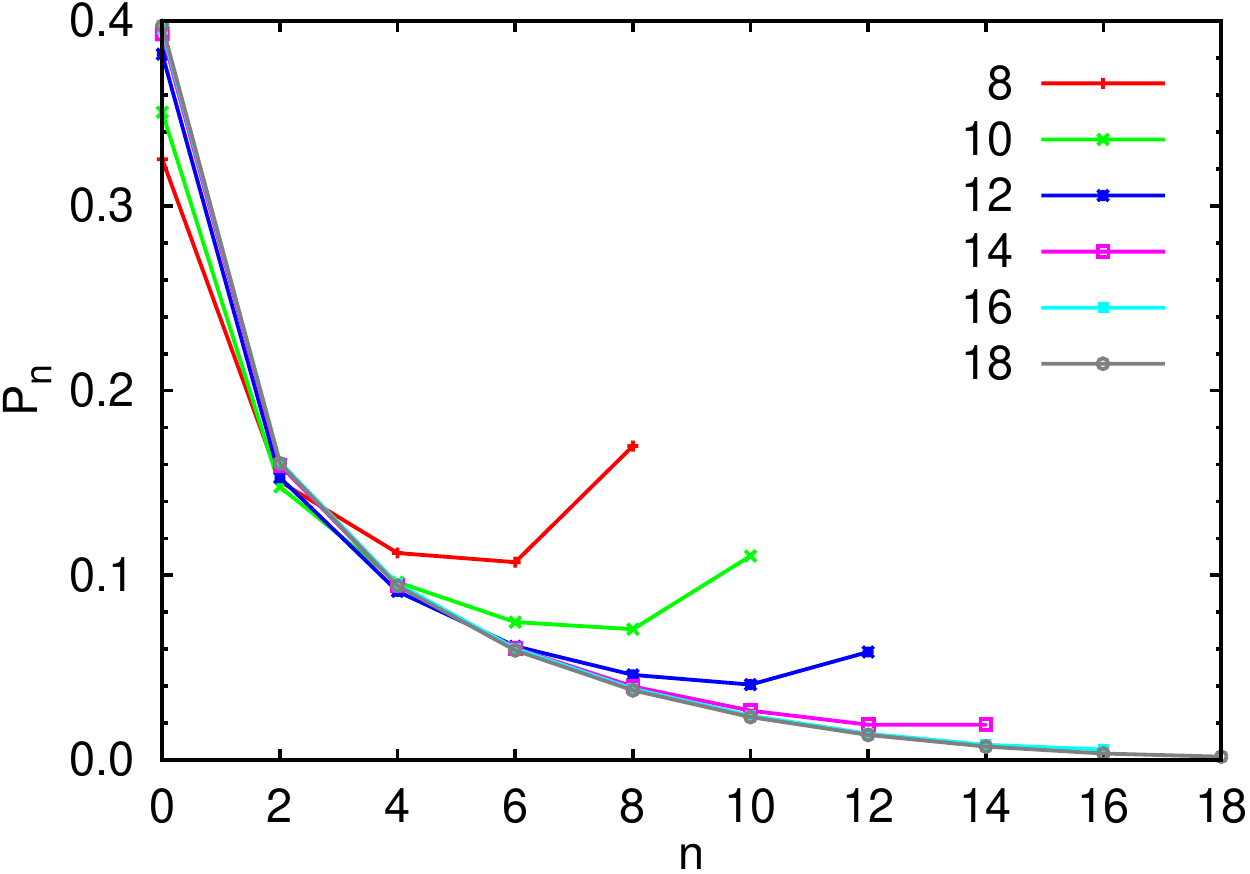}
	\caption{$P_n$ (defined in the caption for Fig. 6) versus $n$ for different number of atoms for $(B=0.001)$. The case $N=16$ and $N=18$ coincide and the result converges for larger $N$. For $N=18$, $\mid \langle PB \mid GS \rangle \mid=0.000$. $A=0.03$ and $gN=6$ has been considered.}
\end{figure}

\begin{figure}[tbp]
	\centering
		\includegraphics*[width=1.0\columnwidth]{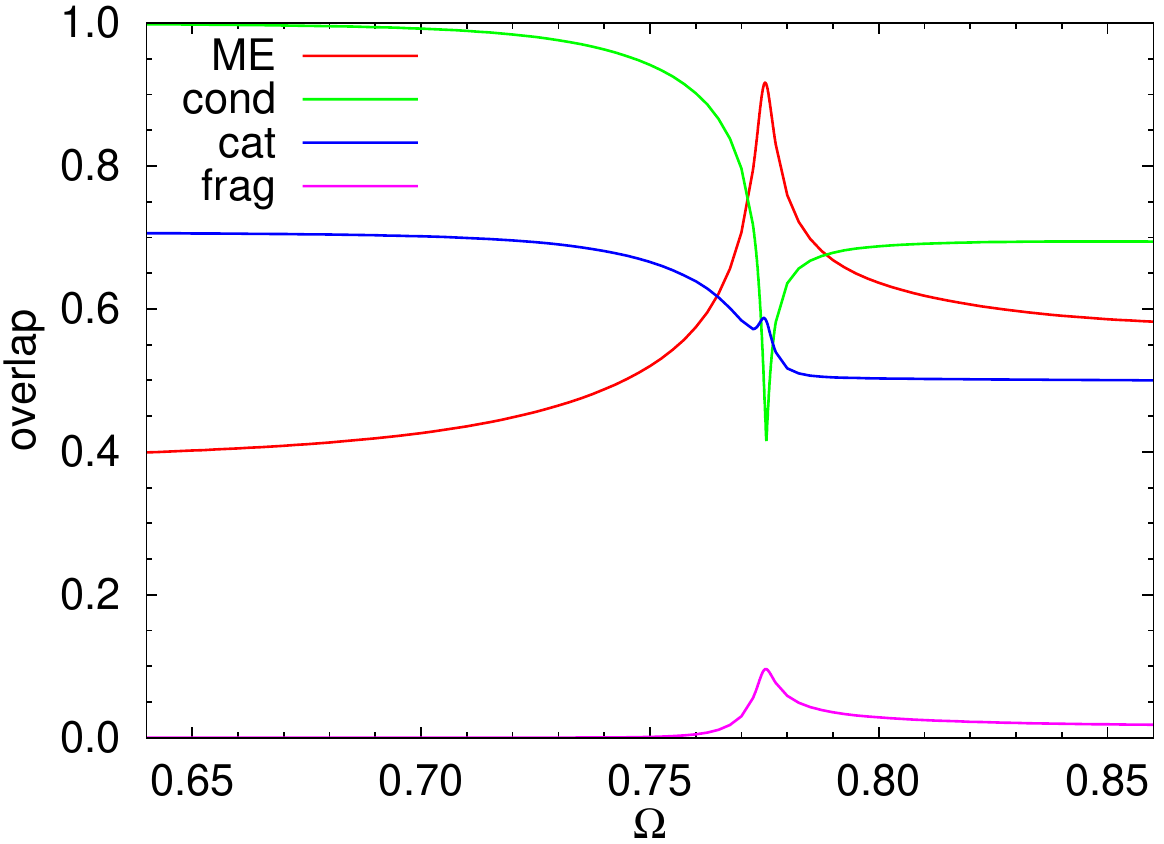}
	\caption{Overlaps, $\mid \langle St \mid GS \rangle \mid$ as a function of $\Omega$, of the GS over different states (St): the maximum entangled (ME), the fully condensed (cond), the cat state (cat) and the fragmented state (frag) for $N=12$, $A=0.03$, $B=0$ and $gN=6$.}
	\end{figure}

To analyze the evolution of the GS in more detail (we focus on the case $B=0$) we look at the eigenfunctions of the SPDM (see Eq. 9), and their occupations $n_k$. They provide an alternative representation of  multiparticle states and operators by the substitution of the FD functions $\varphi_m$ by $\psi_k$. The result obtained is that during the whole time evolution from $\Omega_0$ to $\Omega_f$, only two of these single-particle eigenstates, $\psi_1$ and $\psi_2$ (the most occupied and the next one), with occupations $n_1$ and $n_2$ respectively, play a role in the GS. Moreover, through the whole evolution $(n_1+n_2)/N\geq 0.9$, and as $N$ increases, this joint occupation is even larger. In Fig. 4 the adiabatic change in time of the first two occupations $n_k$ is shown. This result strongly suggests the substitution of the GS by a two-mode state in which only $\psi_1$ and $\psi_2$ are considered. We define as "critical" ($\Omega_c$) the frequency where the coincidence of the occupations $n_1$ and $n_2$ takes place.
 The decrease  of the anisotropy $A$ reduces the width of the critical regime, but the peaks in the population of $\psi_1$ and $\psi_2$ always touch each other. In contrast, if $B\neq 0$, $n_1$ is stricktly greater than $n_2$ at $\Omega_c$, and although these modes are still the dominant ones, 
the peaks do not touch  at any frequency. The two-mode model works well, but  the system becomes slightly more condensed. However, within the region around $\Omega_c$, the system lies always beyond the regime of applicability of MF theory \cite{nature}. In the MF framework, the mechanism that triggers the nucleation at $\Omega_c$ is related with parity symmetry breaking in the order parameter, and this  manifests as dynamical instability, i.e., any slight perturbation of the MF ground state  drives the system far off equilibrium (within the MF dynamics).

Another relevant information, is that only the three first LLL single particle states (for $m=0,1,2$) have a significant weight in the expansion of $\psi_1$ or $\psi_2$. Below $\Omega_c$, $\psi_1$ is a combination of $\varphi_0$ and $\varphi_2$, at $\Omega_c$ it changes its nature to a state that contains only the $\varphi_1$ and remains as that up to $\Omega_f$. The second most occupied state, $\psi_2$, developes the opposite changes in such a way that they interchange their composition at the critical frequency. As a consequence, the density of the GS shows predominantly, two symmetric vortices, produced by the combination of $\varphi_0$ and $\varphi_2$, that move from the edge to the center (up to a non-zero shortest distance), as $\Omega$ approaches $\Omega_c$ and a single centered vortex, produced by $\varphi_1$ from $\Omega_c$ to $\Omega_f$. Fig. 5 shows the evolution of the weights ( $p_0$ and $p_2$) of the FD functions in the expansion of $\psi_1$ up to the critical frequency where they disappear. Close to $\Omega_0$, $\psi_1$ is essentially $\varphi_0$ (and $\psi_1$ is essentially the GS) producing a fully condensed state. Just before $\Omega_c$ it has both components with significant weights and at $\Omega_c$ these weights are substituted by the weight of the unique component $\varphi_1$ (not shown in the figure). 

Had we considered the combination of $\varphi_0$ and $\varphi_2$ for the whole range from $\Omega_0$ to $\Omega$ larger than $\omega_{\perp}$ (with the addition of a quartic potential), in the spirit of a MF approach, then the curves $p_0$ and $p_2$ in Fig. 5 would cross each other, and would end with final values of $p_2=1$ and $p_0=0$ at some $\Omega$, meaning that a single double-quantized vortex would be produced at the centre of the condensate, in agreement with the results obtained by Saito and Ueda \cite{sai}. The evolution would not fulfill the condition of adiabaticity and the last state would be an excited state.

In order to perform simulations of TOF experiments in Section V assuming a large number of particles to have good statistics,  we combine the results from exact diagonalization with the use a two-mode model in the following way: at $\Omega_c$, we analyze the overlap between the exact GS and $N$-body two-mode Fock states of the type, 
\begin{equation}
\ket{n_1,n_2,0,0...}	
\end{equation}
where $n_1+n_2=N$, which will be abbreviated as $\ket{n_1,n_2}$.  Here we must clarify a point concerning the definition of $\psi_1$ and $\psi_2$. At $\Omega_c$ and $B=0$, $n_1=n_2$ and the ``most occupied" state can be both of them. As long as we concentrate in the analysis of the GS at $\Omega_c$, we choose for $\psi_1$ the expansion in terms of $\varphi_0$ and $\varphi_2$ when $B=0$ and for $\psi_2$ the one represented by $\varphi_1$. For a slight $B\neq 0$, $n_1>n_2$, and the previous choice is the right one.

Fig. 6a shows $|<GS|n,N-n>|^2$ as a function of $n$ for $B=0$. The result means that in the GS there is a nearly uniform distribution among the different components, (better as $N$ increases) and in addition, that $\ket{GS}$ is very close to the maximally entangled (ME) state constructed  from even $n$  values, since the overlap $\mid\langle ME\mid GS\rangle\mid = 0.92$ is indeed large, here $\ket{ME}$ is defined as        
\begin{equation}
\ket{ME}=[\ket{N,0}+\ket{N-2,2}+...+\ket{0,N}]/\sqrt{N/2+1}. \label{our}
\end{equation}
Note that we are using here the concept of entanglement for identical particles corresponding to the mode entanglement (for various ways of defining entanglement in systems of identical particles see \cite{eckert,zanardi}.
This analytic approximation to the GS allows us to simulate experimental TOF measurements with an arbitrarily large number of atoms, where the only ingredient supplied by the exact analysis are the coefficients of the expansions of $\psi_1$ and $\psi_2$ on the FD functions. We use it in Section V.
 
Fig. 6b shows the same results for $B\neq 0$. The main difference between the two cases consists on the expansion of $\psi_1$ and $\psi_2$ in the FD states. In the parity broken case, in both expansions, $\varphi_0$, $\varphi_1$ and $\varphi_2$ are significant. The decrease of the columns with $n$ is exponential and can be accurately adjusted by an analytical function. We define as $\ket{PB}$ ($PB$ for parity broken) the expansion on $\ket{n,N-n}$ states with the appropriate decreasing coefficients. It must be pointed out that the single particle odd occupations must be zero    
for $B=0$ and very close to it. The variation of the weights for different number of particles converges as $N$ increases, as shown in Fig. 7 where the results from $N=16$ and $N=18$ coincide. In Section V, we use the converged coefficient to simulate the measurements. For a given $N$ the 
distribution of weights shown in Fig. 6, is robust against changes in the anisotropy strength $A$.
 
For comparison, we also analyzed, for $B=0$,  the overlap of the GS with other celebrated states. 
One is the "cat state" represented by the combination $\ket{N,0}+\ket{0,N}$ \cite{nun}, which means that as a result of a single shot, the system can only appear as a full condensed state of each type with a probability of $50\%$, similar phenomenology has recently been analyzed in Ref. \cite{wri} related with optical vortex cat states. Other is the so called "fragmented state" with only one component in its expansion, given by $\ket{N/2,N/2}$ with a $50\%$ of occupation for each mode \cite{mul},  and the last case considered is the full condensed state. In Fig. 8, for $N=12$, we show the overlaps as a function of $\Omega$ for each case, the system evolves from a full condensed state at $\Omega_0$ to a quite condensed one at $\Omega_f$ passing through a strongly correlated state, very similar to $\ket{ME}$ at the critical frequency.

\begin{figure}[tbp]
	\centering
		\includegraphics*[width=1.0\columnwidth]{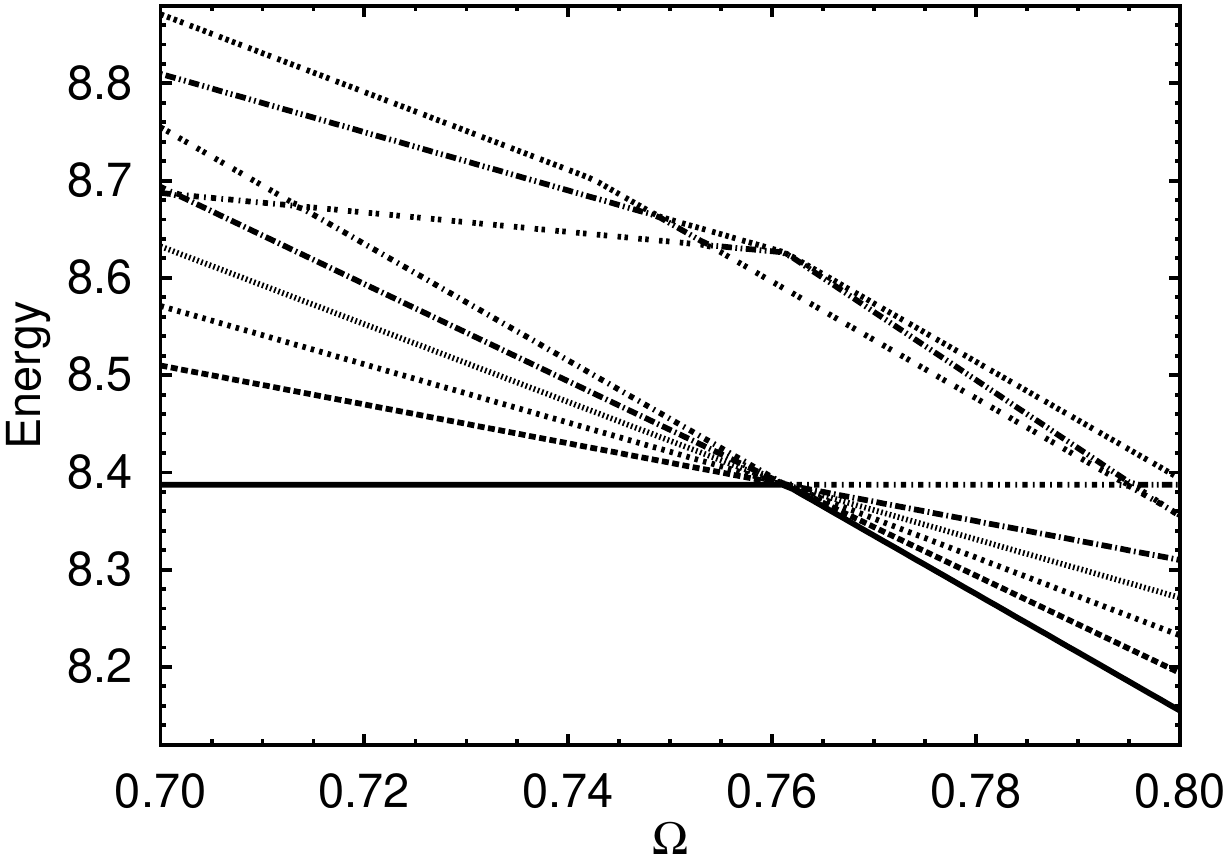}
    \includegraphics*[width=1.0\columnwidth]{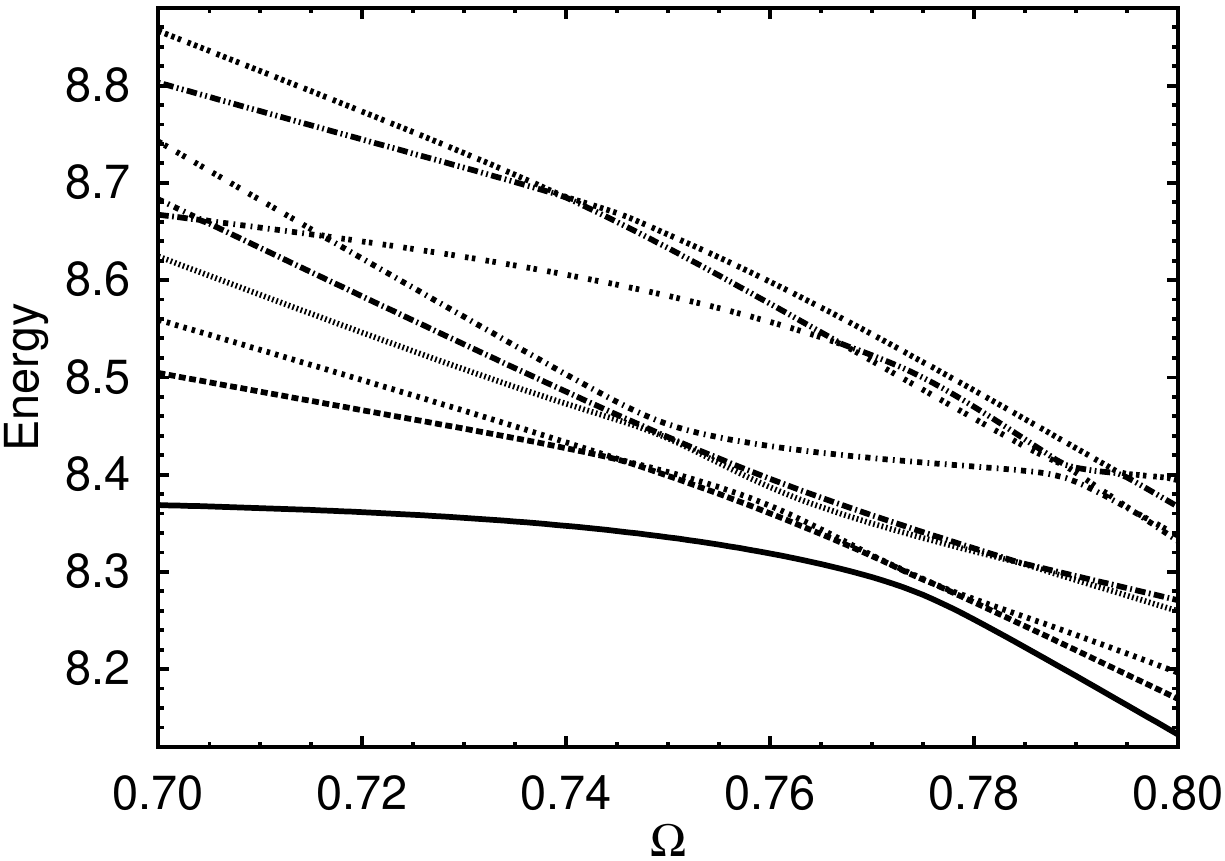}
		\includegraphics*[width=1.0\columnwidth]{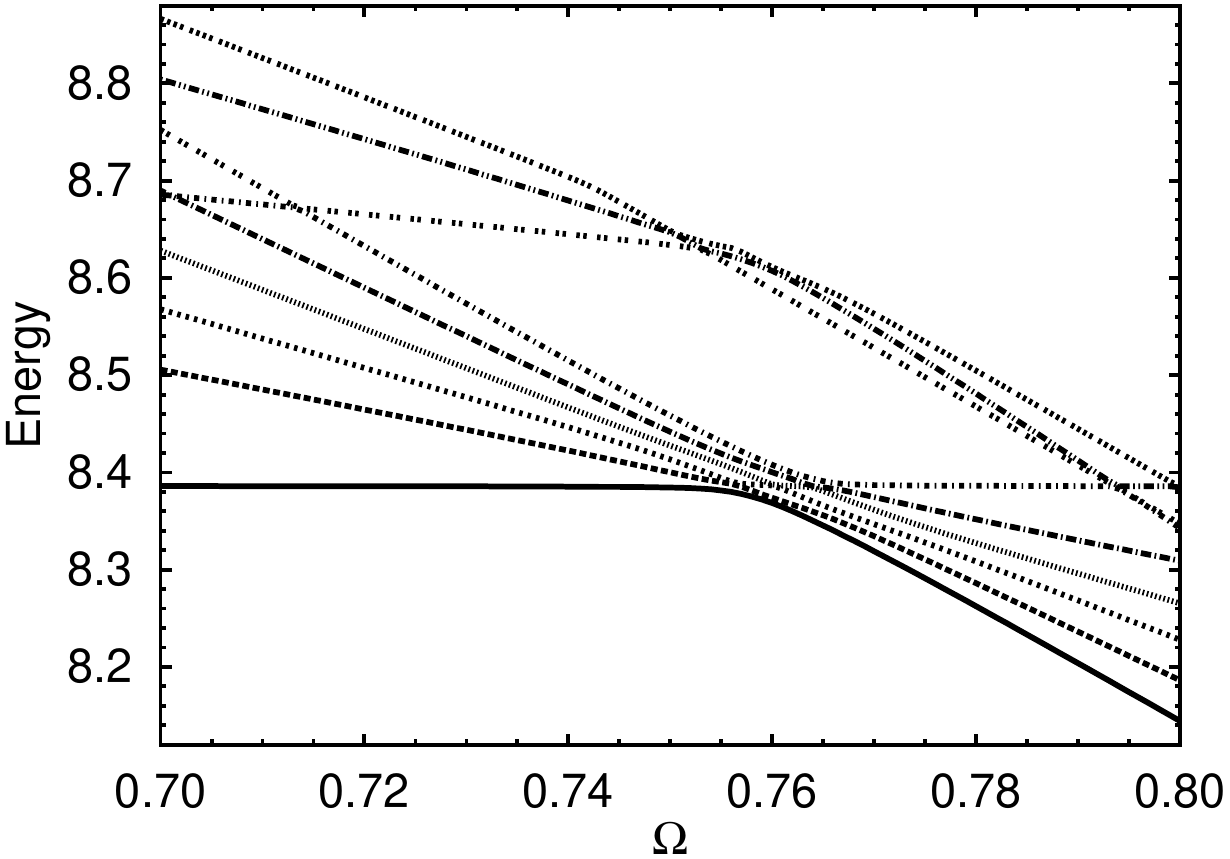}
			
	\caption{ Lowest contributions on the energy spectrum as a function of $\Omega$ for $N=6$ and $gN=6$. In Fig. 8a we consider $A=B=0$, in 8b $A=0.03$ and $B=0$ and in 8c $A=0$ and $B=0.01$ respectively.  The arrows mark the value of $\Omega_1$.}
\end{figure}

\begin{figure}[tbp]
	\centering
		\includegraphics*[width=1.0\columnwidth]{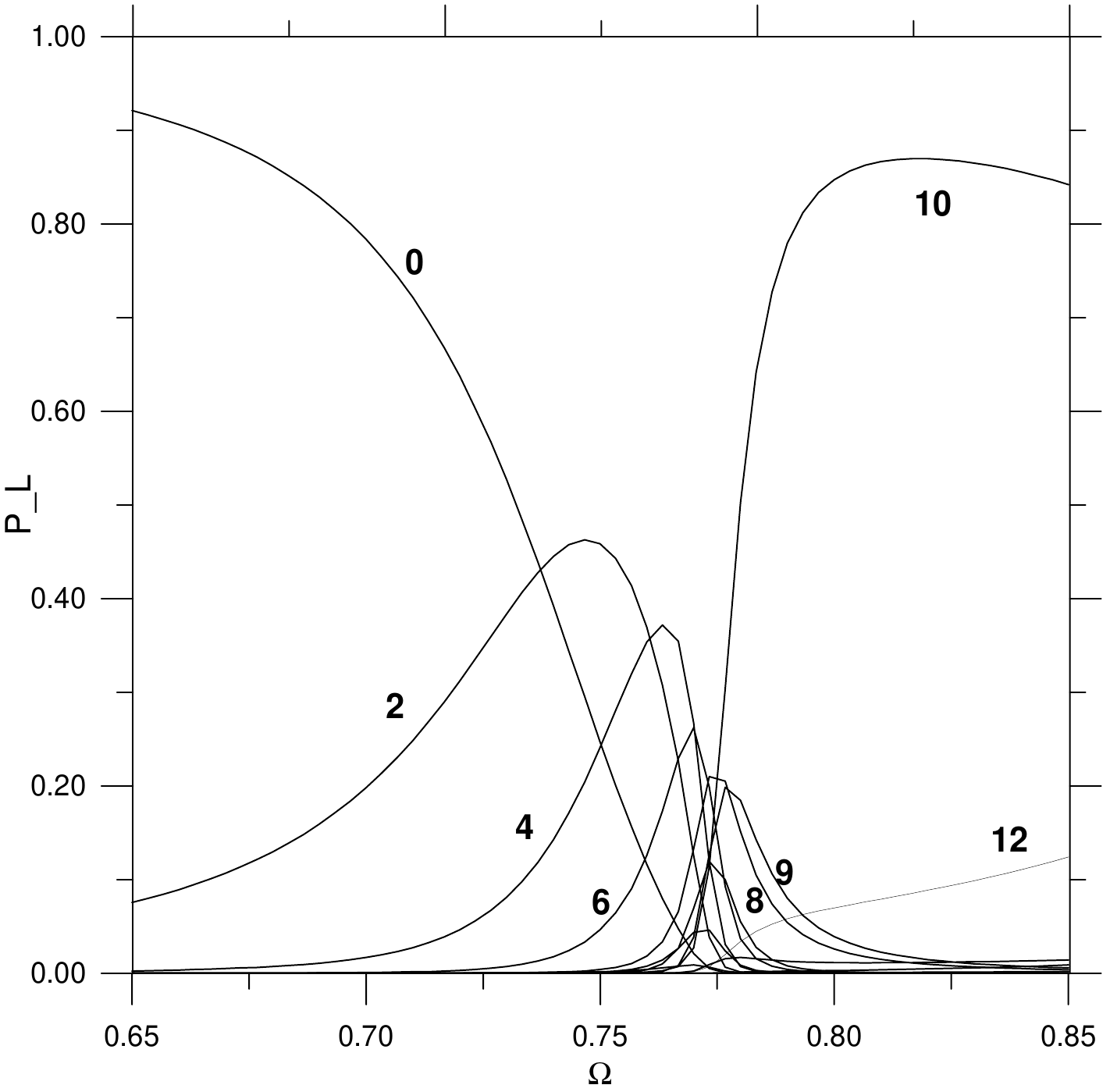}
  	\includegraphics*[width=1.0\columnwidth]{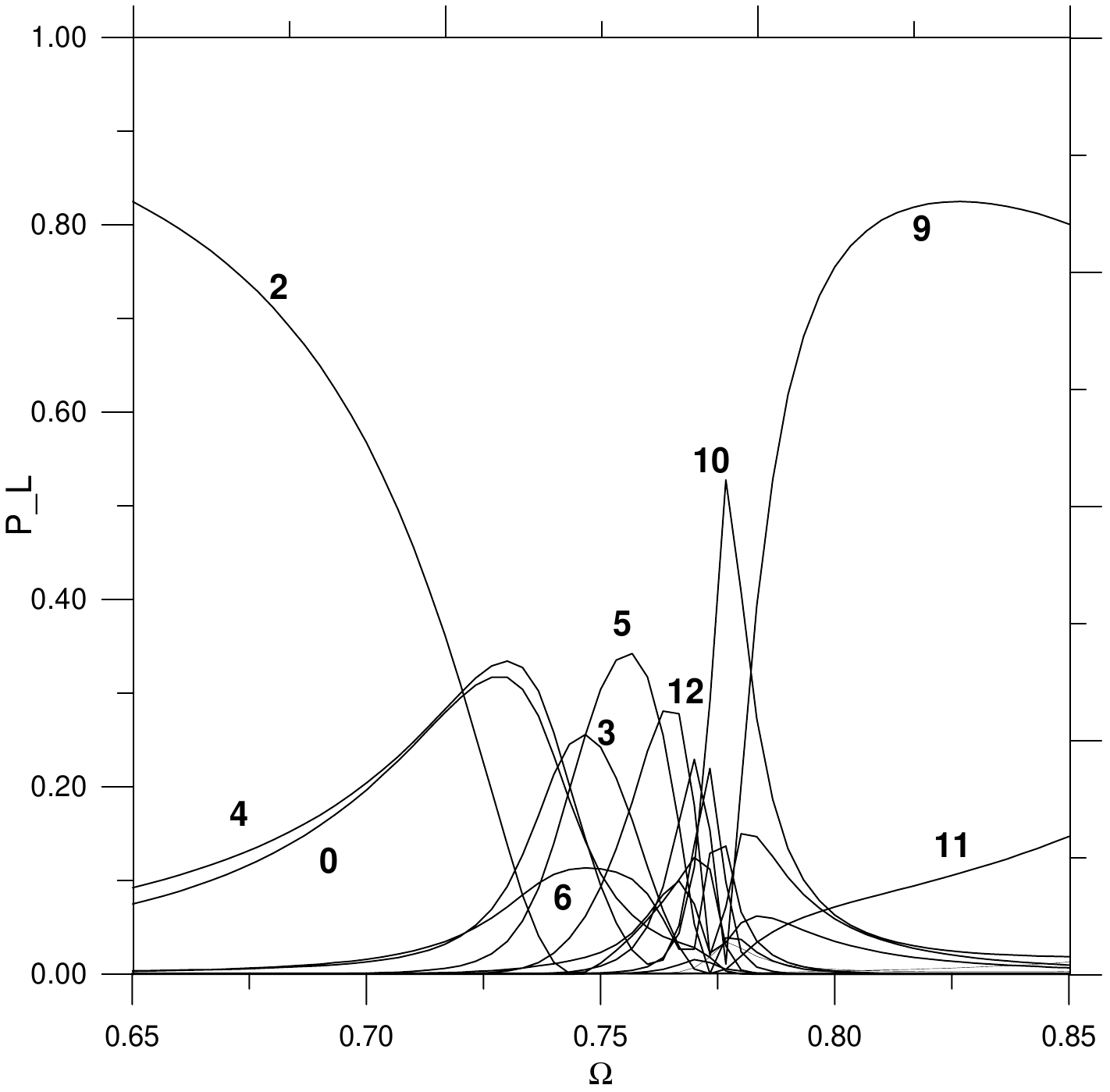}		
	\caption{ Weight of the $L$-subspaces in the composition of the GS (fig. 9a) and the first excited state (Fig. 9b) versus $\Omega$. $N=10$, $gN=6$, $A=0.03$ and $B=0.005$ has been considered.}
	\label{fig:math}
\end{figure}

\begin{figure}[tbp]
	\centering
		\includegraphics*[width=1.0\columnwidth]{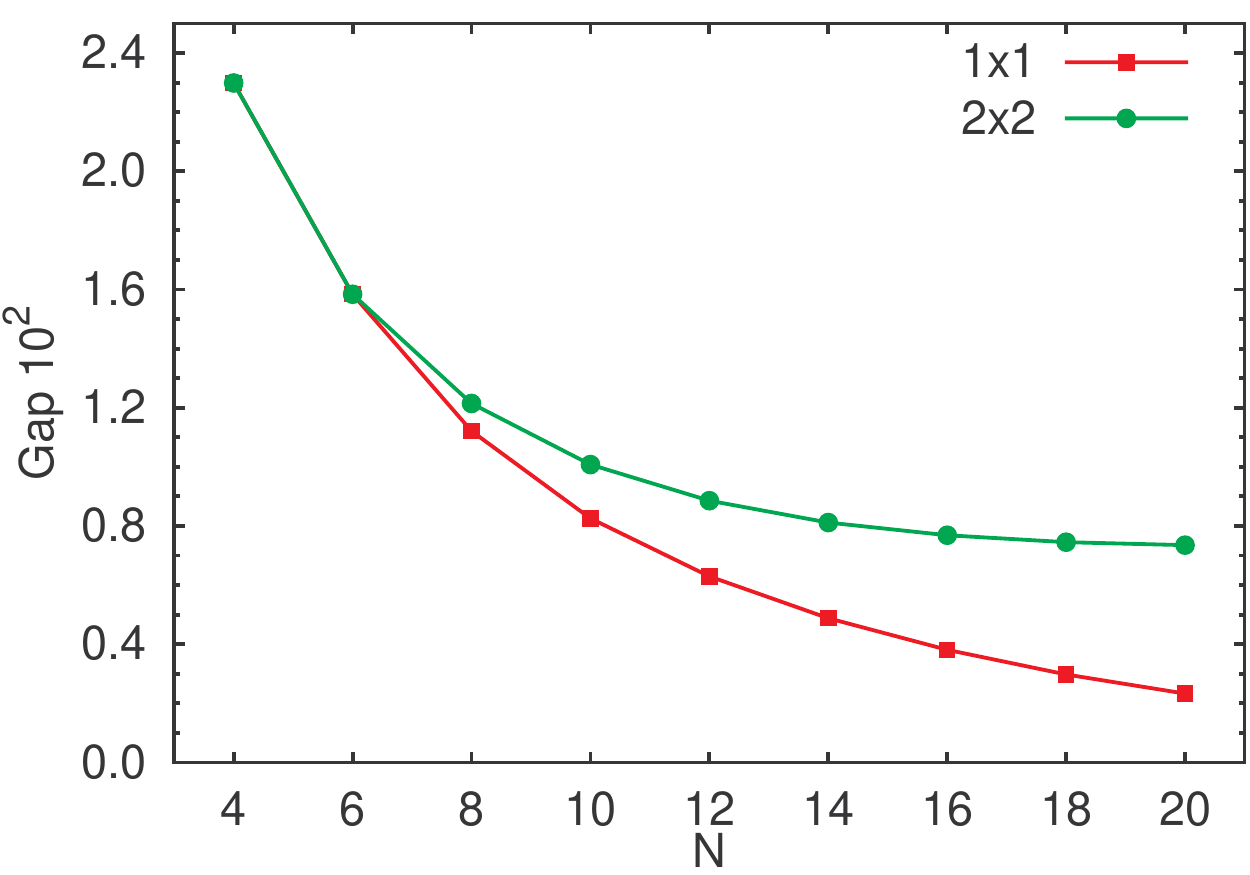}
	\caption{ Minimal gap ($G_m$) between the GS and the first excited state as a function of $N$. In the upper curve only even values of $L$ has been considered ($2\times 2$) , whereas for the lower curve, all $L$ are included ($1\times 1$). $A=0.03$, $B=0$ and $gN=6$ are considered.}  
\end{figure}

\medskip

\section{IV. Energy spectrum}

In Section III we addressed the question related to the possibility of adiabatic evolution from $\Omega_0$ to $\Omega_f$ in a finite $\Delta t$, when non-zero values of $A$ and $B$ are included in the Hamiltonian. The conclusion was that as long as $B$ is small compared with $A$ and $g$,  even for a large number of particles, $\Delta t$ remains finite. In this Section, in contrast, we analyze the energy gap between the GS and the first excited state as a necessary ingredient, aside from the critical $\gamma_c$ already treated in Section III, to decide if the adiabatic criterium is fulfilled. We look for the possibility of excitations of different multipolarity. Firstly, we analyze the energy gap between the GS and the lowest excited states during the evolution in terms of the contribution of the $L$-subspaces in their composition, and see how this analysis depends on $A$, $B$ and $N$. Next we concentrate on the robustness of the GS of the parity-invariant Hamiltonian at the critical frequency, against occasional external perturbations, i.e., we inquire about the experimental feasibility of obtaining the $\Phi_c$ state in the laboratory. This question is suggested by possible applications to quantum information,  where it is necessary to manipulate and control  the system.

For the symmetric Hamiltonian ($A=B=0$), as was mentioned in Section II, at the critical frequency ($\Omega_c=\Omega_1=\omega_{\perp}-gN/8\pi$ in this case), $N$ stationary states are energy degenerated ($L=0$ and $L=2,..,N$) as expected from the analytical expresion of their energies given by,

\begin{eqnarray}
E_L& = & \frac{g}{4\pi}N(N-\frac{L}{2}-1)+(\omega{\perp}-\Omega)L+N\omega_{\perp}
\nonumber
\\
& = &L(\omega_{\perp}-\Omega-\frac{gN}{8\pi})+\frac{gN}{4\pi}(N-1) + N\omega_{\perp} \,\,,
\end{eqnarray}
which becomes $L$ independent when $\Omega=\Omega_1$. This degeneracy is lifted by the introduction of non-zero $A$ or $B$ or both. Fig. 9 shows for $N=10$ the evolution of the energy spectrum for $A=B=0$, $A\neq 0$ and $B=0$, and $A=0$ and $B\neq 0$ respectively. In the second case, the energies are grouped in pairs of different parity, as was previously obtained by Parke et al. (ref), namely, for some values of $\Omega$, the lowest excited state is nearly two-fold degenerated being however, well separated from the GS, for the values of $N$ considered ($N\leq 20$). As $\Omega$ changes, some crossings and anti-crossings take place in such a way that the minimum gap between the GS and the first excitation (defined as $G_m$) may imply jumps from even to even (which is the case of $N\leq 6$), or even to odd (for $N>6$).

In Figure. 10 we show, for $N=10$, the evolution of the contributions of different $L$-subspaces in the GS (Figure 10a) and in the first excited state (Figure 10b). We consider $A=0.03$, and a relatively large value of $B=0.005$ to emphasize the broken-parity effect. Since both symmetries are broken, all $L$-subspaces have non-zero contributions, however there is a remarkable difference between the regions below and above $\Omega_1=0.76$. For $\Omega<\Omega_1$ only even values of $L$ are significant in spite of the fact that the Hamiltonian breaks parity symmetry,  and the presence of external fluctuations that would produce monopolar ($L\pm 1$) or octupolar ($L\pm 3$) excitations is irrelevant, since these excitations are energetically blocked; the most probable process is a quadrupolar excitation from $L=0$ to $L=2$. In contrast, for $\Omega>\Omega_1$ all even and odd values of $L$ play a role; in particular,  far from $\Omega_1$, the change from $L=10$ to $L=9$ in the most probable scenario, which means that the possible  non-adiabaticity of the evolution would be dominated by a braking process (i.e., lowering the speed), where in an effective way, one of the atoms jumps from the condensed to the thermal phase, ceasing its contribution to the total angular momentum of the system. Had we suppressed $B$ in Fig. 10, we would  obtain for the first excited state, separated regions with only even or only odd contributions (due to crossings in the spectrum) and in particular at the critical frequency, the jump would be from even to even for $N\leq 6$ and even to odd for $N>6$.             

Finally, in Fig. 11 we show the minimal gap $G_m$ as a function of $N$ obtained from the parity-invariant Hamiltonian. We distinguish between two cases: in the upper curve ($2\times\ 2$) only even values of $L$ are considered, i.e., is the gap for quadrupolar excitations, in contrast, in the lower branch all $L$-values are considered ($1\times 1$) and represents the minimal gap that must be overcome by any perturbation of arbitrary multipolarity. If the system in protected against parity-breaking perturbations, the adiabaticity of the process is guaranteed (upper curve) since $G_m$ tends to a constant for large $N$ and simultaneously $\gamma_c$ decreases, in such a way that the adiabatic criterium given by $\gamma_c N/G_m \ll 1$ is fulfilled. However, if parity breaking excitations can occur and the number of particles is large, the adiabatic evolution is practically impossible. 

It is worth noticing that we have been dealing with two different definitions of the critical frequency $\Omega_c$, one is the frequency where the two single particle occupations equalize ($n_1=n_2$), and the other one is the frequency where the minimum gap between the GS and the first excited state takes place; within our numerical precision both definitions are the same.  

\begin{figure}[tbp]
	\centering
		\includegraphics*[width=1.0\columnwidth]{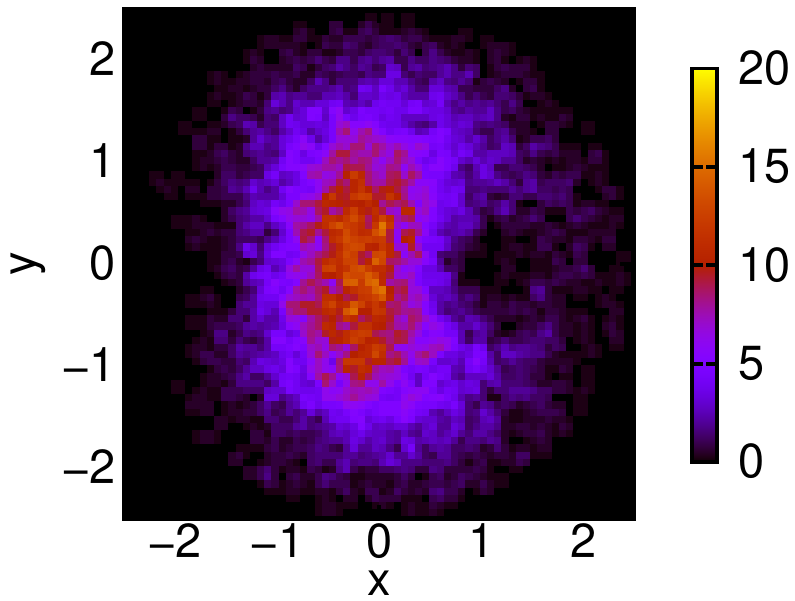}
    \includegraphics*[width=1.0\columnwidth]{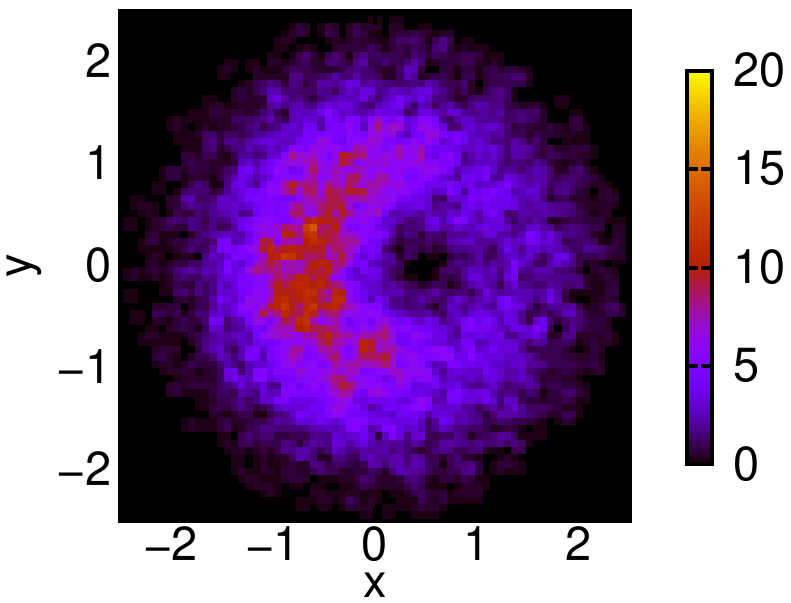}
		\includegraphics*[width=1.0\columnwidth]{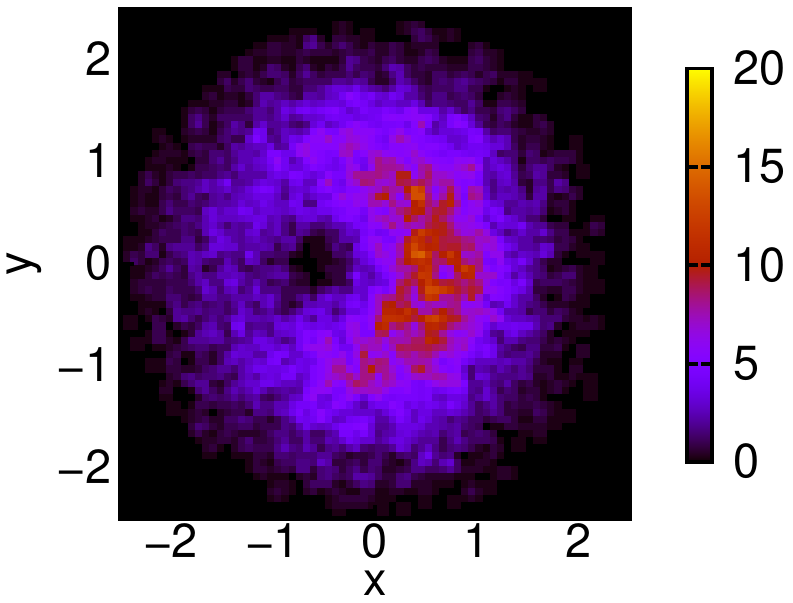}
   \includegraphics*[width=1.0\columnwidth]{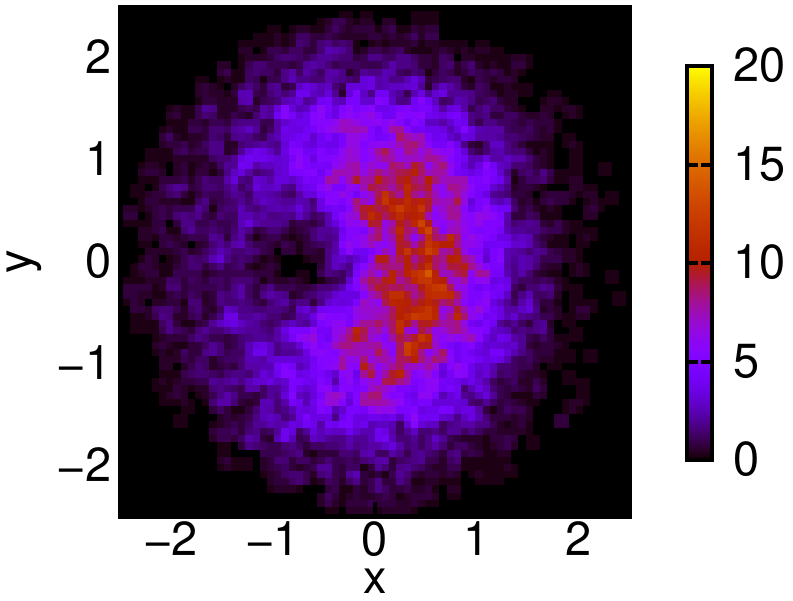}	
	\caption{ Set of 4 single shots on the $\ket{ME}$ state, for $N=10000$, $A=0.03$ and $B=0$.}  
\end{figure}

\begin{figure}[tbp]
	\centering
		\includegraphics*[width=1.0\columnwidth]{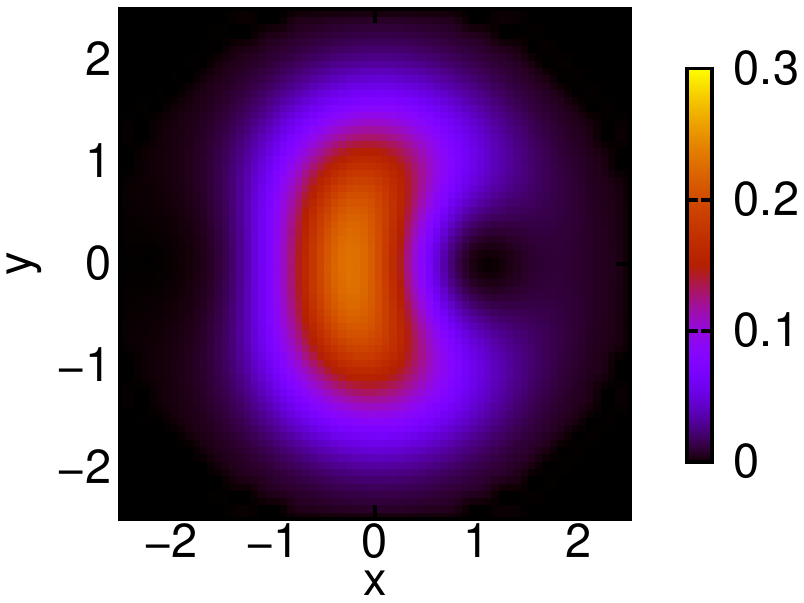}
   \includegraphics*[width=1.0\columnwidth]{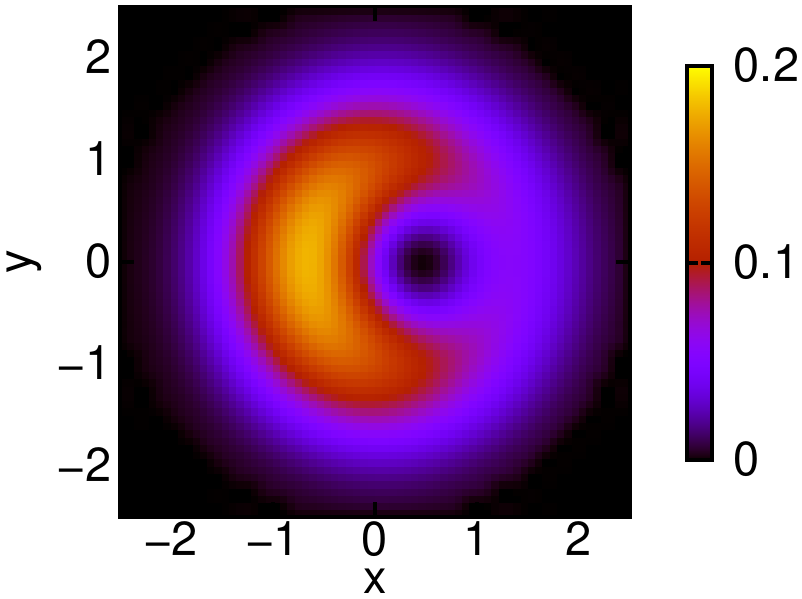}
		\includegraphics*[width=1.0\columnwidth]{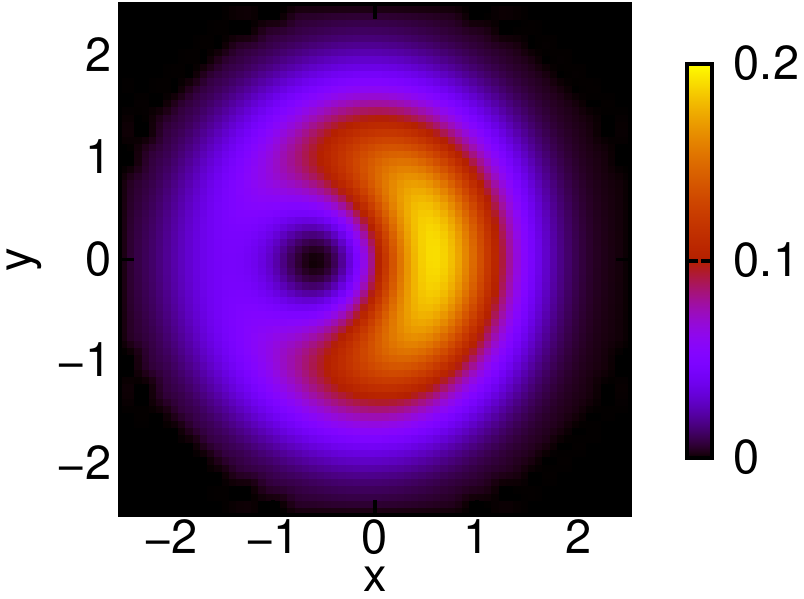}
   \includegraphics*[width=1.0\columnwidth]{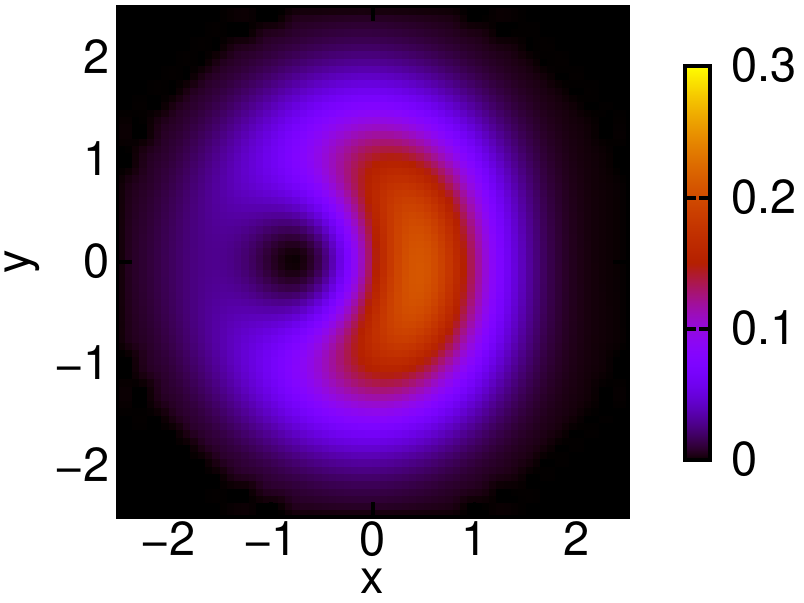}	
	\caption{ N-th correlation functions of the shots of Figure 12 see Eq.19.}  
\end{figure}

\begin{figure}[tbp]
	\centering
		\includegraphics*[width=1.0\columnwidth]{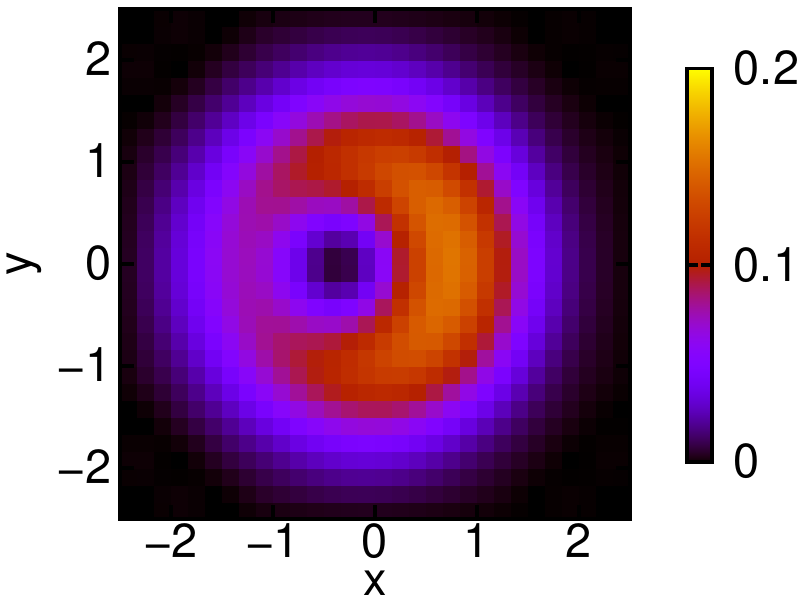}
		\includegraphics*[width=1.0\columnwidth]{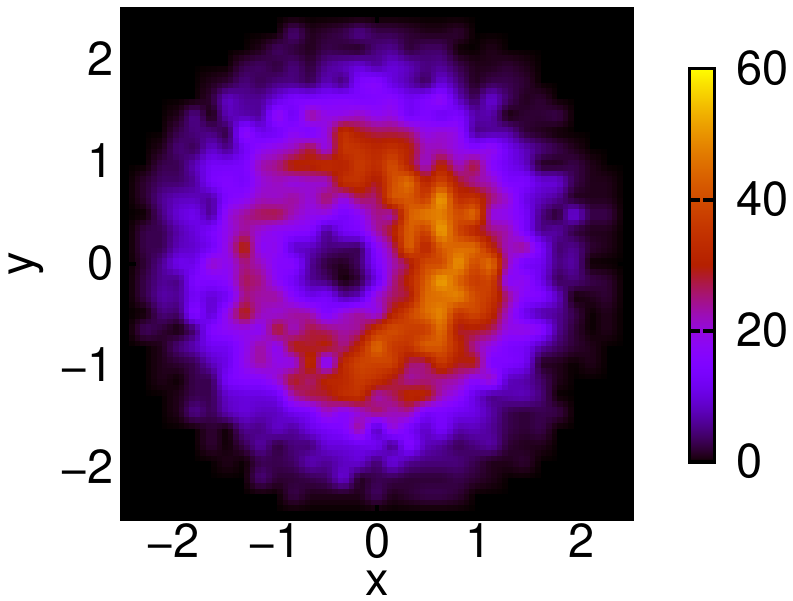}
	\caption{ A single shot on the $\ket{PB}$  state (PB for parity-broken) and its N-th correlation function for $N=1000$, $A=0.03$ and $B=0.001$.  }  
\end{figure}

\begin{figure}[tbp]
	\centering
		\includegraphics*[width=1.0\columnwidth]{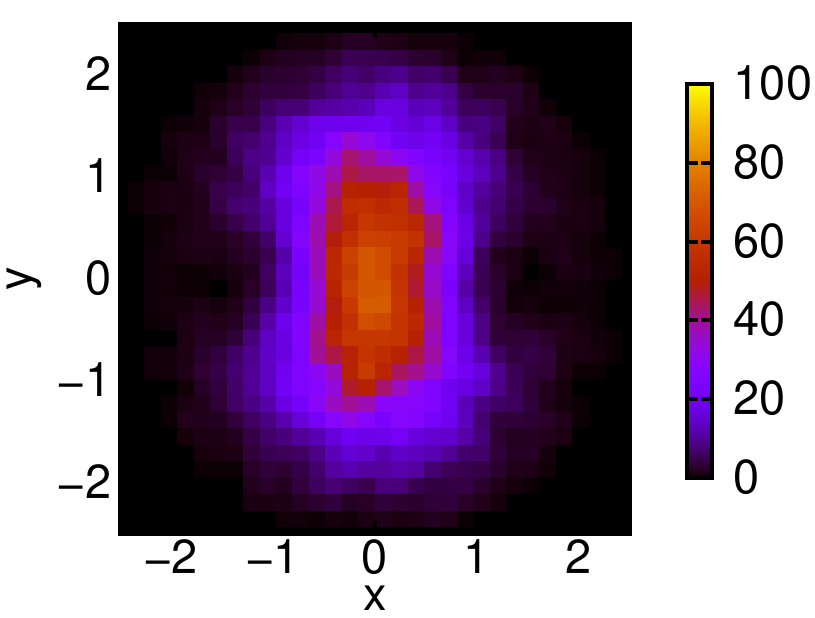}		
    \includegraphics*[width=1.0\columnwidth]{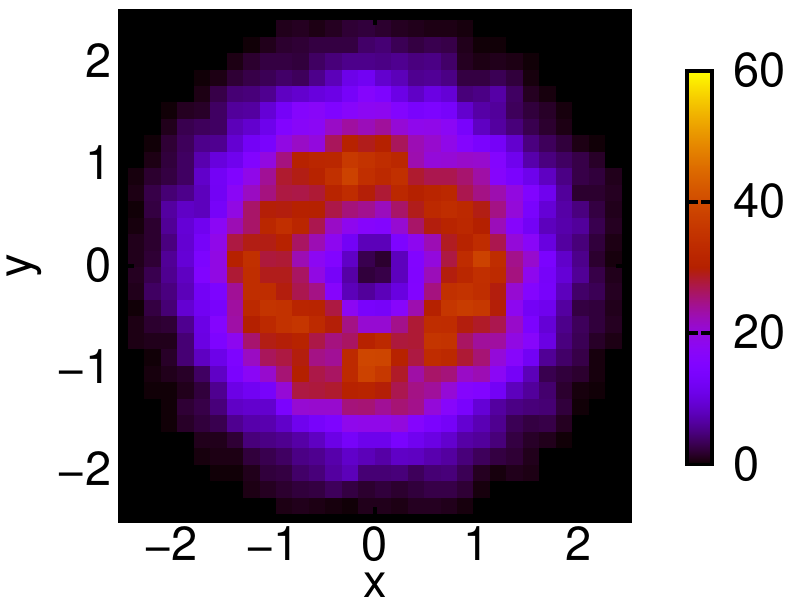}				
	\caption{Shots on the Cat State.}  
\end{figure}

\medskip
\section{V. Simulation of measurement}

Once we have an accurate representation of the GS at the critical frequency for both cases, with and without parity symmetry breaking, we can simulate measurements during a TOF experiment for an arbitrary large number of atoms \cite{jav}. We assume balistic expansion, a hypothesis compatible with our LLL condition \cite{rea} . We proceed as follows: after solving the eigenvalue equation for the SPDM of the exact GS (see Eqs. 9 and 10), we get $\psi_1$ and $\psi_2$ that define the two modes. Using Eq. 17 we obtain two different realizations of the two-mode model, defined as $\ket{P0}$ for $B=0$ and $\ket{PB}$ (PB for parity-broken) for $B\neq 0$, both of them approximate the GS. Then the density distribution of this GS determines the position of the first atom using the following algorithm. A randomly generated position $\vec{r}$ is accepted if $\rho(\vec{r})/\rho_{max}$ is larger than another randomly generated number $u$, $0\leq u \leq 1$ or rejected otherwise; $\rho_{max}$ being the maximum of the density. This first step ends up with the detected position of the first atom, let's call it $\vec{r}_1$. Next we consider the pair correlation function (PCF) given by $\rho^{(2)}(\vec{r}_1,\vec{r})=<GS\mid \hat{\Psi}^\dag (\vec{r_1})\hat{\Psi}^\dag (\vec{r}) \hat{\Psi}(\vec{r})\hat{\Psi}(\vec{r_1})\mid GS>$  ( being $\hat{\Psi}(\vec{r})=\sum_l \varphi_l \hat{a}_l\,\,$ the field operator ) and generate the second position in the same way. By repetition of the procedure $N$-times, finally we get the set of positions of all the $N$ atoms. This sequential algorithm simulates a TOF measurement of a single shot. The procedure is the same in both cases, with or without parity broken symmetry however, the main difference is that in the first case ($B\neq 0$), the weights of the components $\ket{n,N-n}$ in $\ket{PB}$ become negilible for $n>18$, which simplifies the numerical calculation. 

Fig. 12 shows a set of 4 shots for $N=10000$ atoms with $B=0$; here the spots represent the positions of the $N$ atoms, whereas in Fig. 13 we show the $N$-correlation function defined as,
\begin{equation}
\rho^{(N)}(\vec{r}_1,\vec{r}_2,...,\vec{r}_{N-1},\vec{r})
\end{equation}
related to the $4$ shots of Fig.12 respectively. The $N$-correlation gives the probability to find the last atom at $\vec{r}$, once the other $N-1$ have fixed positions. For such a large number of atoms we obtain, as a general result, that the $j$-correlation function is indistinguishable from the $j+1$ one, if $j\geq J\sim 100$, namely the correlation is important for the first positions, but ceases to be modified by the addition of more fixed positions, due to screening effects, somehow this number $J$ is related with the locality of the correlation. From the sample of 4 single shots shown in Fig. 12, it is clear that a vortex is produced at random places along the $X$-axis (this axis is preferred due to the particular form of the anisotropic term considered), with sometimes a slight manifestation of a second vortex, as is the case of the first shot. At odds with this result, in the case where parity symmetry has been broken (i.e., the last degree of freedom), the GS is projected onto the most probable option and the vortex always appears at the same place (see Fig. 14), at a negative value of $x$ in our case. This result is again due to our particular choose of the parity-broken term in the Hamiltonian, namely, the vortex would be located at a positive $x$ if
\begin{equation}
V_2(x)=-B\frac{M\omega_{\perp}^2}{\lambda_{\perp}}\sum_i^N x_i^3
\end{equation}
had been chosen. 

We interpret the results in the following way: each realization of the system in a single shot (in the case $B=0$ for example) is produced by two ingredients, one is the ``intrinsic'' nature of the GS which determines the density distributions of the $j\leq N$ successive correlation functions, (or equivalently the density of systems with $N-j+1$ particles), and the other is the particular measurement procedure, in our case by a one-by-one detection of the atoms. In other words, the measurement modifies the system and the pictures shown are a combination of  the two factors. Different type of experiments detecting particle positions, would produce different pictures, coming however from the same GS. However, the differences disappear in the averaged picture given by the density, the experimental mechanism does not modify the mean properties of the system.

According to the result demonstrated in Appendix B, the accumulation of a large number of shots for a macroscopic system, reproduces the density. However, a non expected result is that the density of a system of $N=10000$ particles is indistinguishable from that of a reduced $N$ ($N\leq 20$), meaning that some mean properties of macroscopic systems are well captured by the exact results from mesoscopic systems making unnecessary the extrapolation to the thermodinamic limit. This density contains two vortices symmetricaly positioned along the $x$-axis. We want to stress that at the level of the exact GS we proved that the holes that appear in the density are real single quantized vortices as the phase of $\psi_1$ changes by $2\pi$ around them. If the analysis of the nucleation process would had only the density of the sequence of stationary states for increasing $\Omega$ from $\Omega_0$ to $\Omega_c$, as a unique source of information, the conclusion would be that the nucleation of the first vortex is preceded by the presence of two symmetrically  positioned vortices that move to the center. However, this is nothing else than a possibility within the multiple realizations of the experimental performance.

As a test of our procedure to simulate the TOF experiment, we considered a cat state, artificially created from the $\ket{P0}$ supressing all the contributions shown in Fig. 6a except those from $\ket{0,N}$ and $\ket{N,0}$. The result is shown in Fig. 15, which corresponds to the densities of $\psi_1$ and $\psi_2$ with one or two vortices as expected. We obtained only two possibilities, as the system is fully occupying the single particle $\psi_1$ or $\psi_2$ and no partial occupations are possible. The positions of the non-centered vortices is given by $x = (\sqrt{2}\mid\frac{\alpha}{\beta}\mid)^{1/2} \sim 1.5$ where $\psi_1=\alpha \varphi_0+\beta \varphi_2$, in terms of the FD states.      

Some experience in the analysis of many-body systems suggests that complementary information can be obtained by the inspection of the PCF \cite{roma}. However, the success of this excercise may be strongly different for different type of states. Since the meaning is the probability of finding an atom at $\vec{r}$ when another one is placed at $\vec{r}_1$, a circular symmetric result means that the system is not correlated, whereas a hole at $\vec{r}_1$ and ordered peaks in special positions, means that the position of the atoms is strongly correlated; this is, for example,  the case of the Laughlin state, solution of the symmetric Hamiltonian with $L=N(N-1)$ in the region of strong rotation.  This is a non degenerated GS, and as a consequence, its dendity is circular symmetric preserving the symmetry of the Hamiltonian. However, its PCF reveals a strongly correlated Wigner type structure of $N$ peaks. The reason for the discrepancy between the density and the PCF is that the Laughlin state in a number state with the phase completely undefined (the opposite case of a condensate which is a phase state), it is a linear combination of all possible orientations of the Wigner structure. A measurement that fixes the position of one atom projects the system in a particular orientation, revealing the Wigner structure. In contrast, the GS at $\Omega_c$ is the solution of a Hamiltonian that contains a rotating symmetry-breaking term ($A\neq 0$) and has a fixed orientation. The PCF does not provide then extra unknown information if $\vec{r_1}$ is chosen at the maximum of the density.

Finally, we want to mention a speculation. In the case of a system submitted to both types of anisotropy ($A\neq 0$ and $B\neq 0$), and considering in general both signs for the parameter $B$, the density of the system would contain two vortices in fixed symmetric positions on the $x$-axis. This state is a candidate to experience tunneling between the two single vortex states in agreement with the picture raised by Parke {\it et al.} \cite{parke} as the precursor mechanism for the nucleation.

\section{VI. Conclusions}

We have  analyzed in this paper  vortex nucleation in mezoscopic 2D Bose superfluid in a rotating trap.
The main ingredient of our work is that we have  included a weakly anisotropic stirring potential, breaking thus explicitly the axial (rotational) symmetry. The system we consider is well described by the mean field theory well below "criticality" (with an even  condensate wave function), and  well above the "criticality", with the order parameter being the  odd function. This phenomenon involves therefore a discrete parity symmetry breaking. In the critical region the MF solutions exhibit dynamical instability. The main result of our paper is that the true many body state is a strongly correlated entangled state involving two macroscopically occupied modes (eigenvectors of the single particle density matrix). We have characterize this state in various aspects, which can be summarized in more details as follows: 

\begin{itemize}

\item The parity symmetry breaking at the critical frequency manifests itself as dynamical instability within the mean field framework. It  does not prevent, however,  the adiabaticity of the nucleation process. The increase of the parity conserving perturbation $A$ notably reduces the necessary period of time to evolve from $\Omega_0$ to $\Omega_f$ (for $A=0.1$, $\Delta t \sim 1$ s.). This conclusion remains valid  even when the number of particles increases. However, a significant value of the parity breaking perturbation of $B$ evidently acts  against the adiabaticity. This perturbation leads to an exponential decrease of the energy gap from the GS to the first excited state. 

\item The maximally entangled combination of $\psi_1$ and $\psi_2$ of the two mode state, which is a fairly accurate representation of the strongly correlated state at the critical frequency for $B=0$, reveals a single vortex structure randomly located along the $x$-axis in a single shot measurement (with an additional small probability of a pair of, in general, non symmetrically located  vortices). This is the result of the particular way of measurement  mechanism, that we consider here; a one-particle-followed-by-another-one detection.

\item The function $\rho^{(2)}(\vec{r}_1,\vec{r})$ with $\vec{r}_1$ at one of the peaks of the density on the $y$-axis, does not reveal any hidden structure, due to the fact that the system has a fixed orientation, and the position of the vortex along the $x$-axis is smeared out by the integration over the positions of the other $N-2$ atoms. This is an intrinsic property of  this correlation function. In contrast, $\rho^{N}(\vec{r})$ breaks both symmetries, rotational and parity, producing the pictures shown in Fig.13 typical of a projection mechanism implicit in a single measure \cite{mul}.
  
\item The state $\ket{ME}$ becomes a better representation of the exact GS as $N$ increases. It is robust against changes in $A$ for $0\leq A \leq 0.1$. The state $\ket{PB}$ has zero contribution of $\ket{n,N-n}$ for $n\geq 18$.

\item The mean properties of the system as those given by the total energy and possible by the density along the evolution in the whole range of variation of $\Omega$ considered, are insensitive to the symmetry broken mechanisme at $\Omega_c$. 

\item Instability or chaotic behaviour of the system in a mean field calculation can be a signature of the existence of a strongly correlated state which description lies beyond the mean field framework.

\end{itemize}

\vskip1cm
\section{Appendix A}

\paragraph{Natural orbitals}
From the diagonalization of the SPDM, one obtains a new set of single particle wave functions (spwf), often refered as natural orbitals, and their corresponding creation and annihilation operators,
\begin{eqnarray}
	\psi_i &=& \sum_{j=0}^L\beta_{ij}\varphi^{FD}_j\\
	\hat{b}_i &=& \sum_{j=0}^L\beta_{ij}\hat{a}_j
	\label{eq:bet}
\end{eqnarray}
where $\beta_{ij}$ are real numbers. $\hat{a}_j$ is the operator that annihilates the FD with angular momentum $j$, $j= 0,1,...,L$ being $L$ the largest single particle angular momentum in the GS $\ket{\Psi_0}$. We have sorted the spwf's in decreasing order of occupation ($\ave{n_i}$), in such a way that $\hat{b_1}$ and $\hat{b_2}$ create the most occupied single particle and the next one respectively. It is worth to notice that the subindex in $\hat{b}_i$ and $\psi_i$ ($i=1,2,..., L+1$) labels different SPDM eigenstates, whereas the subindex in $\varphi^{FD}_i,\,$ ($i=0,1,...,L$) means angular momentum.\\

The representation of a state $\ket{\Psi}$ in terms of the FD functions can be transformed into one in terms of the natural base $\{ \psi_i\}$ in the following way:

Given a general state
\begin{equation}
	\ket{\Psi}=\sum_{k=1}^{dim}\alpha_k\ket{k}_{FD}
	\label{eq:Psi}
\end{equation}
where $\ket{k}_{FD}$ is a $N$-body state expressed in the FD base and $dim$ is the dimension of the space considered, $\ket{k}_{FD}$ can be expressed as,
\begin{equation}
	\ket{k}_{FD}=\frac{1}{\sqrt{\prod_{l=0}^L n_{lk}!}}\prod_{j=1}^N\hat{a}^\dagger_{w_{kj}}\ket{0}
	\label{eq:c1}
\end{equation}
where $n_{lk}$ is the occupation of FD $l$ in the state $\ket{k}_{FD}$, and ${w_{kj}}$ is the angular momentum of each particle in the state $\ket{k}_{FD}$. From eq. \ref{eq:bet} and eq. \ref{eq:c1} we obtain for eq. \ref{eq:Psi},
\begin{equation}
	\ket{\Psi}=\sum_{k=1}^{dim}\frac{\alpha_k}{\sqrt{\prod_{l=0}^L n_{lk}!}}\prod_{j=1}^N
	\sum_{i=1}^{L+1}\beta_{i,w_{kj}}
	\hat{b}^\dagger_{i}\ket{0},
	\label{eq:c2}
\end{equation}
where we have used the ortogonality properties of the $\beta$ matrix: $\{\beta_{ij}\}^{-1}=\{\beta_{ij}\}^T=\{\beta_{ji}\}$.\\

\paragraph{Overlap}
The ME state is a combination of states $\ket{n,N-n,0,...}_{SPDM}$, $n\in even$, with the same weight, where $n$ is the occupation of the most occupied spwf and $N-n$ the occupation of the next one. At $\Omega_c$ the first natural orbit is a combination of spwf with angular momentuma $m=0$ and $m=2$, $\varphi^{SPDM}_1= c\,\varphi^{FD}_0+ d\,\varphi^{FD}_2$, with  $d=-\sqrt{1-c^2}$ and the second one is equal to the FD with angular momentum $m=1$, $\varphi^{SPDM}_2= \varphi^{FD}_1$. Then, each state can be expressed in the FD bases as
\begin{eqnarray}
 &&\ket{n,N-n,0,...,0}_{SPDM}= \frac{(c\hat a_0^\dagger+d\hat a_2^\dagger)^n(\hat a_1^\dagger)^{N-n}}{\sqrt{n!(N-n)!}} \ket{0}= \nonumber\\
 &&\sum_{i=0}^n\bin{n}{i}\frac{(c\hat a_0^\dagger)^{n-i}(d\hat a_2^\dagger)^i}{\sqrt{n!}} \ket{0,N-n,0,...,0}_{FD}=  \nonumber\\
 &&\sum_{i=0}^n\sqrt{\bin{n}{i}}\,c^{n-i}\,d^i\ket{n-i,N-n,i,0,...,0}_{FD}
 \label{st_ME}
\end{eqnarray}
Projecting each of this states on the GS obtained from exact diagonalization $\ket{GS}=\sum_{i_0,i_1,...,i_L}\alpha_{i_0,i_1,...,i_L}\ket{i_0,i_1,...,i_L}_{FD}$, summing over all the states $\ket{n,N-n,0,...}_{SPDM}$, $n\in even$ and finally multiplying by the normalization constant $1/\sqrt{N/2+1}$, we obtain the overlap expressed in the simple form
\begin{eqnarray}
 &&\abs{\bk{GS}{ME}}= \frac{1}{\sqrt{N/2+1}}\left| \sum_{\substack{n=0\\n\in even}}^N\sum_{i=0}^n \sqrt{\bin{n}{i}}\cdot \right. \nonumber\\
 &&c^{n-i}\,d^i\,\alpha_{n-i,N-n,i,0,...,0} \Bigg|
 \label{over_ME}
\end{eqnarray}
which measures the suitability of $\ket{ME}$ as an approximation of the GS. 

\section{Appendix B}

\paragraph{The density}
The superposition of the data coming from a large number of single shots reproduces the density of a system in a state $\ket{\Psi}$, as can be shown in what follows.

The probability to find a particle at the position $\mathbf{r}$ after $k$ particles have been detected, is given by the function ($k\leq N-1$, $N$ the number of particles),
\begin{eqnarray}
	& & \rho^{(k+1)}(\mathbf{r_1},\mathbf{r_2},\cdots,\mathbf{r_k},\mathbf{r}) = \left\langle\, \cfield{\mathbf{r_1}} \cfield{\mathbf{r_2}} \cdots \cfield{\mathbf{r_k}} \cdot \right. \nonumber \\
	& & \left. \cfield{\mathbf{r}} \field{\mathbf{r}} \field{\mathbf{r_k}} \cdots \field{\mathbf{r_2}} \field{\mathbf{r_1}} \,\right\rangle
\end{eqnarray}
where the expected value is in the state $\Psi$ and  $\field{\mathbf{r_k}}=\sum_i\varphi_i(\mathbf{r_k})\hat a_i$ is the field operator. Using the commutation relations of the creation and annihilation operators and the ortogonormalization of the set $\{\varphi_i(\mathbf{r_k})\}$ we can deduce the general relations
\begin{widetext}
\begin{eqnarray}
\int_{\mathbf{r_1}\in\mathds{R}^2} \rho^{(2)}(\mathbf{r_1},\mathbf{r})d\mathbf{r_1} &=& (N-1)\,\,\rho(\mathbf{r}) \label{eq:cor1} \\
\int_{\mathbf{r_1}\in\mathds{R}^2} \int_{\mathbf{r_2}\in\mathds{R}^2} \rho^{(3)}(\mathbf{r_1},\mathbf{r_2},\mathbf{r})d\mathbf{r_1}d\mathbf{r_2} &=&  (N-2)(N-1)\,\,\rho(\mathbf{r}) \\
& \vdots & \\
\int_{\mathbf{r_1}\in\mathds{R}^2} \int_{\mathbf{r_2}\in\mathds{R}^2} \ldots \int_{\mathbf{r_{N-1}}\in\mathds{R}^2} \rho^{(N)}(\mathbf{r_1},\mathbf{r_2},\cdots,\mathbf{r_{N-1}},\mathbf{r})d\mathbf{r_1}d\mathbf{r_2}\cdots d\mathbf{r_{N-1}} &=&  (N-1)!\,\,\rho(\mathbf{r})\,\,.
\end{eqnarray}
\end{widetext}
We could use any of them to recover the density, however the last is the one that fits our simulation. To model the experiment we have defined a grid in the $xy$-plane and we count the number of times that we detect a particle at each site of the grid. On the other hand, if we interprete the multiple integral of Eq. 12 as a multiple summation on a large number of different configurations $\{\mathbf{r_1},\mathbf{r_2},...,\mathbf{r_{N-1}},\mathbf{r}\}$ on the discretized grid, keeping $\mathbf{r}$ fixed, then we can make the connection since the histogram obtained after a large number of shots, is nothing else than, aside of a constant number, the probability to find a particle at $\mathbf{r}$ when the other $N-1$ have visited all the possible configurations, which is the meaning of the correlation function $\rho^{(N)}$ in the left hand side of Eq. 12. More than that, the summation over the $\rho^{(N)}$ functions is not arbitrary, it contains the information of the structure of the state $\Psi$ as is the case in the simulation. It can be easily proved for example, in the case of the pair correlation function that can be rewritten as
\begin{eqnarray}
\rho^{(2)}(\mathbf{r_1},\mathbf{r}) & = & \left\langle V_0\mid \cfield{\mathbf{r_1}}\cfield{\mathbf{r}}\field{\mathbf{r}}\field{\mathbf{r_1}}\mid V_0 \right\rangle \nonumber
\\
& = & \rho(\mathbf{r_1})\left\langle V_1\mid \cfield{\mathbf{r}}\field{\mathbf{r}}\mid V_1 \right\rangle
\end{eqnarray}
where
\begin{equation}
\ket{V_1}=\frac{\field{\mathbf{r_1}}\ket{V_0}}{\sqrt{\rho( \mathbf{r_1})}}
\end{equation}
is a system with $N-1$ particles and $\ket{V_0}$ is the initial state $\ket{\Psi}$. Or in other words, the probability to find a particle at $\mathbf{r}$ when other is located at $\mathbf{r_1}$ depends on the probability to have a particle at $\mathbf{r_1}$ in the state $\Psi$. This complets our assertion.

\section{Acknowledgements}

We acknowledge very fruitful collaborations with Jean Dalibard, whose contribution to this project is comparable to those of the authors. 

We acknowledge Spanish MEC/MINCIN projects TOQATA (FIS2008-00784 and FIS2007-60350) and QOIT (Consolider Ingenio 2010). M.L. acknowledges ESF/MEC project FERMIX (FIS2007-29996-E), EU Integrated Project  SCALA, EU STREP project NAMEQUAM, ERC Advanced Grant QUAGATUA, and Alexander von Humboldt Foundation Senior Research Prize.

\end{document}